# A Multiscale Constitutive Model for Metal Forming of Dual Phase Titanium Alloys by Incorporating Inherent Deformation and Failure Mechanisms

Umair Bin Asim[1, 2], M Amir Siddiq[1,*], Robert M McMeeking[1, 3, 4], Mehmet E Kartal[1]

[1] School of Engineering, University of Aberdeen, Fraser Noble Building, AB24 3UE, Aberdeen, United Kingdom

[2] Department of Materials Science and Engineering, Texas A&M University, College Station, Texas 77843, USA

[3] Materials Department, University of California, Santa Barbara, CA 93106, USA

[4] Department of Mechanical Engineering, University of California, Santa Barbara, CA 93106, USA

E-mail: amir.siddiq@abdn.ac.uk

## Abstract

Ductile metals undergo a considerable amount of plastic deformation before failure. Void nucleation, growth and coalescence is the mechanism of failure in such metals. $\alpha - \beta$ titanium alloys are ductile in nature and are widely used for their unique set of properties like specific strength, fracture toughness, corrosion resistance and resistance to fatigue failures. Voids in these alloys were reported to nucleate on the phase boundaries between $\alpha$ and $\beta$ phase. Based on the findings of crystal plasticity finite element method (CPFEM) based investigation of the void growth at the interface of $\alpha$ and $\beta$ phases [1], [2], a void nucleation, growth, and coalescence model has been formulated. An existing single-phase crystal plasticity theory is extended to incorporate underlying physical mechanisms of deformation and failure in dual phase titanium alloys. Effects of various factors (stress triaxiality, Lode parameter, deformation state (equivalent strain), and phase boundary inclination) on void nucleation, growth and coalescence are used to formulate the constitutive model while their interaction with a conventional crystal plasticity theory is established. An extensive parametric assessment of the model is carried out to quantify and understand the effects of the material parameters on the overall material response. Performance of the proposed model is then assessed and verified by comparing the results of the proposed model with the RVE study results. Application of the constitutive model for utilisation in the design and optimisation of the forming process of $\alpha - \beta$ titanium alloy components is also demonstrated using experimental data.

## 1. Introduction

Dual-phase $\alpha - \beta$ titanium alloys are used to a greater extent than any other titanium alloy. The unique properties and mechanical behaviour of dual-phase $\alpha - \beta$ alloys, including specific strength, high and low cycle fatigue resistance, ductility, toughness, corrosion resistance make them an ideal candidate for the applications in areas ranging from aerospace, automotive, energy, and oil & gas sectors. $\alpha - \beta$ titanium alloys have $\alpha$ and $\beta$ phases present in the microstructure in various morphologies which undergo considerable plastic deformation before failure by the ductile mechanism. Their deformation behaviour has been linked and found to be greatly influenced by the underlying microstructure [3]–[5]. Furthermore, void nucleation, growth and coalescence has been reported as the failure mechanism in dual phase $\alpha$ - $\beta$ alloys [6], and it was also identified that voids nucleate on $\alpha-\beta$ phase boundaries [7]. Such phenomena were studied by Asim *et al.* [1], [2] utilizing a crystal plasticity finite element method (CPFEM) approach based on a representative volume element. It is a purpose of the current paper to present a homogenised constitutive model of the Gurson-Tvergaard-Needleman (GTN) type that encompasses not only the elastoplastic response of dual phase $\alpha$ - $\beta$ titanium alloy, but also the special nature of void nucleation, growth and coalescence at phase boundaries in such alloys that leads to their ductile failure. We also verify, validate and characterise the constitutive model, and illustrate its application to sheet metal forming, including the forming limit associated with ductile necking and fracture.

A brief but all-encompassing review of prior efforts to develop constitutive models that can predict the elastic-plastic deformation behaviour of $\alpha-\beta$ titanium alloys is presented in the following.

CPFEM based studies to understand the effect of microstructure morphology for dual phase titanium have been performed in the past [1], [8]–[15] using a representative volume element based approach (for details see ref [1], [2] and references therein). We note that most of the research to formulate constitutive models of $\alpha-\beta$ titanium alloys is focused on high temperature elasto-plastic deformation [16]–[27]. The constitutive models for this alloy can be classified into those that are empirical or semi empirical models [16]–[19], [23], [24] and those that are physics based [20]–[22], [25]–[27]. Empirical or semi empirical models have the advantage of lesser number of parameters. However such models cannot capture microstructural evolution and deformation mechanisms at the crystalline level. The relevant research [16]–[27] has focussed on predicting work hardening [16]–[20], [22], [23], [27], dynamic recrystallisation [16], [21], [24]–[26], and ductile damage [21]. Most of these works were focussed on uniaxial tension or compression simulations and did not consider loading complexities, such as the effects of stress triaxiality and the Lode parameter, that are associated with sheet metal forming processes [1], [2], [28], [29].

In the context of analytical and numerical modelling of ductile behaviour of general porous metallic materials, a detailed description of the state of the art has been recounted in earlier work [2] and is summarised here to justify the novelty of the current paper. The pioneering works of Rice & Tracey [30] and Gurson[31] are the backbone of almost all subsequent models [9], [23], [29], [32]–[53]. Some recent and relevant works are discussed briefly below and the reader is directed to the references therein for further details and history. Stewart & Cazacu [36] developed a macroscopic yield criterion including the effects of material anisotropy, incompressibility and tension-compression asymmetry. A macroscale homogenisation based model, accounting for the stress triaxiality and the Lode parameter to understand evolution of void shape and orientation, was presented by Danas and Ponte Castañeda [38]. The localisation of plastic flow around a void in a rate-independent, isotropic, elastic-plastic Levy-von Mises material was studied by Dunand and Mohr [40] under different stress states considering shear loading and the Lode parameter. Similar studies were carried out by Tekoğlu et al. [41] and Torki and Benzerga [42]. Zhou et al. [9] enhanced the GTN model by combining volumetric and shear damage in low-stress triaxialities. Song and Ponte Castañeda [43], [44] presented a macroscale homogenisation-based constitutive model for a porous, single phase, polycrystalline material, while Niordson and Tvergaard [46] developed a macroscale constitutive model for porous materials based on strain gradient plasticity theory. In addition, Siddiq [29] incorporated the effects of various parameters including initial porosity, stress triaxiality and crystal orientation on void growth and failure in a porous crystal plasticity constitutive model for single phase FCC material using the results obtained from an extensive RVE study of the same material [54].

It can be seen that previous works have concentrated mostly on single phase materials with the majority of the constitutive models being at the macroscale. Such models do not consider void nucleation and growth mechanisms inside the grains and at the grain/phase boundaries and hence cannot be used to accurately predict failure in single- or bi-crystals. Also, these models do not take into account the combination of various effects that can be important, such as state of deformation (equivalent strain), stress triaxiality, Lode parameter and phase boundary inclination) on the void growth at the interface of the $\alpha$ and $\beta$ phases of dual phase titanium alloys.

As noted above, it is a purpose of the current paper to present a comprehensive homogenised constitutive model for a dual phase $\alpha$ - $\beta$ titanium alloy. Based on work carried out earlier on the modelling of dual phase titanium alloy [1], [2], it was established that a constitutive model for this material needs to take into account anisotropy, other relevant microstructural features, and the effects of deformation state (equivalent strain), stress triaxiality, Lode parameter and phase boundary inclination with particular reference to void nucleation, growth and coalescence at the interface of the $\alpha$ and $\beta$ phases of the dual phase titanium alloy. In the previous studies of void growth at the interface of dual-phase $\alpha$-$\beta$ titanium alloy [1], [2], a non-porous crystal plasticity model was calibrated for single crystals of each of the two phases ($\alpha$ and $\beta$) using experimental data from the literature. In addition, the research [1], [2] showed that the terminal mechanism for dual phase titanium alloys is ductile failure, and must be incorporated in the constitutive model to make it effective for the design and optimisation of forming processes which are used for the fabrication of parts and components. In the present paper, results from these studies [1], [2] are used to extend the non-porous single phase crystal plasticity model to include the effects of various factors including stress triaxiality, the Lode parameter, deformation state (equivalent strain), and phase boundary inclination (PBI) on void nucleation, growth and coalescence in dual phase titanium alloys. The resulting constitutive model is able to capture elastoplastic deformation and failure, and so can be used to predict performance of the alloy in processes such as sheet metal forming. As noted above, in this paper we verify, validate and characterise the constitutive model, and illustrate its application to sheet metal forming including failure at the forming limit.

The article is organised as follows. In Section 2, we present the crystal plasticity based constitutive model which accounts for void nucleation, growth and coalescence. In Section 3, we present a parametric assessment of the proposed model. The model verification and validation is presented in Section 4 followed by constitutive model application in the context of forming limit curve simulations in Section 5. We finally present the conclusions in Section 6. The numerical implementation of the proposed model is described in detail in Appendices.

# 2. Constitutive model

## 2.1 Updated kinematics

The total deformation gradient is multiplicatively split into elastic and plastic parts and the elastic part is further split into an elastic stretch, $\boldsymbol{V}^e$ and a rigid body rotation, $\boldsymbol{R}^e$ using the following relations:

$$\boldsymbol{F} = \boldsymbol{F}^e \boldsymbol{F}^p; \qquad \boldsymbol{F} = \boldsymbol{V}^e \boldsymbol{R}^e \boldsymbol{F}^p \tag{2-1}$$

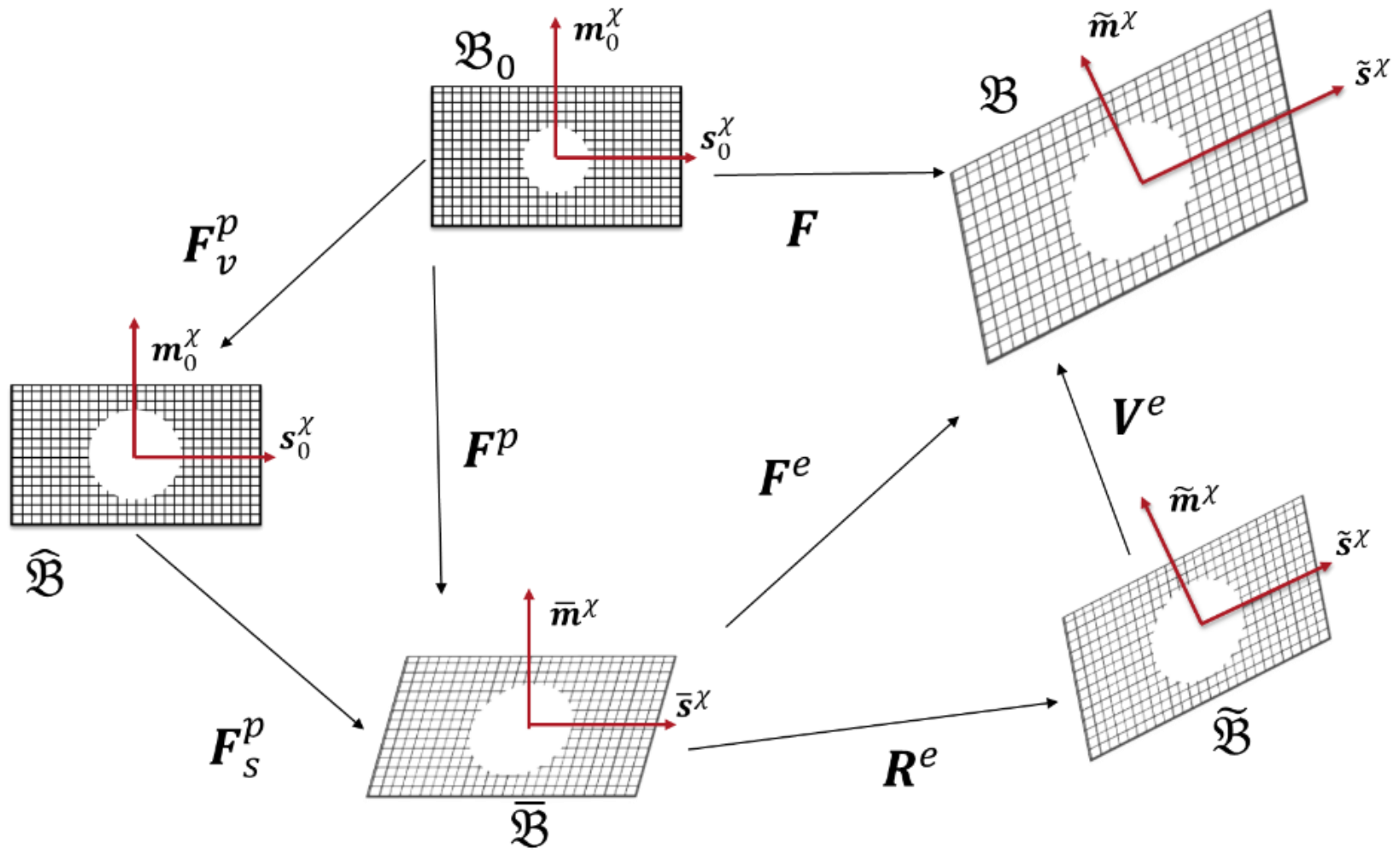

*Figure 2-1: The multiplicative decomposition of total deformation gradient*

Figure 2-1 shows the decomposition of the total deformation gradient. The rigid body rotation and the plastic part of the deformation gradient can be combined to form an unloaded intermediate configuration given by:

$$\boldsymbol{F} = \boldsymbol{V}^e \boldsymbol{F}^*;\ \boldsymbol{F}^* = \boldsymbol{R}^e \boldsymbol{F}^p \tag{2-2}$$

The plastic deformation gradient can further be split into plastic deformation due to slip, $\boldsymbol{F}_s^p$, and plastic deformation due to void growth, $\boldsymbol{F}_v^p$, using:

$$\boldsymbol{F}^p = \boldsymbol{F}_s^p \boldsymbol{F}_v^p \tag{2-3}$$

The total velocity gradient, $\boldsymbol{l}$, is given by:

$$\boldsymbol{l} = \dot{\boldsymbol{F}}\boldsymbol{F}^{-1} = \dot{\boldsymbol{V}}^e \boldsymbol{V}^{e^{-1}} + \boldsymbol{V}^e \tilde{\boldsymbol{L}}^* \boldsymbol{V}^{e^{-1}} \tag{2-4}$$

Here, $\tilde{\boldsymbol{L}}^*$ is velocity gradient in the intermediate unloaded configuration which can be related to the plastic velocity gradient, $\bar{\boldsymbol{L}}^p$, using:

$$\tilde{\boldsymbol{L}}^* = \dot{\boldsymbol{F}}^* \boldsymbol{F}^{*^{-1}} = \dot{\boldsymbol{R}}^e \boldsymbol{R}^{e^T} + \boldsymbol{R}^e \bar{\boldsymbol{L}}^p \boldsymbol{R}^{e^T} \tag{2-5}$$

The plastic velocity gradient can then be divided into the slip and the void growth parts using the following relation:

$$\bar{\boldsymbol{L}}^p = \dot{\boldsymbol{F}}^p \boldsymbol{F}^{p^{-1}} = \dot{\boldsymbol{F}}_s^p \boldsymbol{F}_s^{p^{-1}} + \boldsymbol{F}_s^p \hat{\boldsymbol{L}}_v^p \boldsymbol{F}_s^{p^{-1}} \tag{2-6}$$

Here, $\hat{\boldsymbol{L}}_v^p$ is the velocity gradient due to void growth. The velocity gradient due to slip, $\bar{\boldsymbol{L}}_s^p$, can be related to shear strain rate due to slip, $\dot{\gamma}^\chi$, using:

$$\bar{\boldsymbol{L}}_s^p = \dot{\boldsymbol{F}}_s^p \boldsymbol{F}_s^{p^{-1}} = \sum_{\chi=1}^{N} \dot{\gamma}^\chi\, \bar{\boldsymbol{s}}^\chi \otimes \bar{\boldsymbol{m}}^\chi \tag{2-7}$$

Here, $\bar{\boldsymbol{s}}^\chi$ and $\bar{\boldsymbol{m}}^\chi$ are the unit vectors along the slip direction and normal to the slip plane in the crystal coordinate system and $\otimes$ is the outer or dyadic product. The velocity gradient due to void growth is related to the rate of change of a non-dimensional strain like quantity, $\xi$, which is the amount of void growth, using the following relation:

$$\hat{\boldsymbol{L}}_v^p = \frac{1}{3} A_n \dot{\xi} \boldsymbol{1} \tag{2-8}$$

Here, $A_n$ is a material parameter and $\boldsymbol{1}$ is the second-order identity tensor.

The total plastic velocity gradient can then be defined by:

$$\bar{\boldsymbol{L}}^p = \sum_{\chi=1}^{\mathrm{N}} \dot{\gamma}^\chi\, \bar{\boldsymbol{s}}^\chi \otimes \bar{\boldsymbol{m}}^\chi + \boldsymbol{F}_s^p \frac{1}{3} A_n \dot{\xi} \mathbf{1} \boldsymbol{F}_s^{p^{-1}} \tag{2-9}$$

Since the void growth part of the velocity gradient only deals with volume change, transformation of the volume part using $\boldsymbol{F}_s^p$ will have no effect and the total plastic velocity gradient will then be:

$$\bar{\boldsymbol{L}}^p = \sum_{\chi=1}^{\mathrm{N}} \dot{\gamma}^\chi\, \bar{\boldsymbol{s}}^\chi \otimes \bar{\boldsymbol{m}}^\chi + \frac{1}{3} A_n \dot{\xi} \mathbf{1} \tag{2-10}$$

The relationship for the total velocity gradient given in (2-4) can be rearranged to give the velocity gradient in the intermediate configuration and can then be additively decomposed into the pure elastic stretch, the rigid body rotation and the plastic deformation as:

$$\tilde{\boldsymbol{L}} = \boldsymbol{V}^{e^{-1}} \boldsymbol{l} \boldsymbol{V}^e = \boldsymbol{V}^{e^{-1}} \dot{\boldsymbol{V}}^e + \tilde{\boldsymbol{L}}^* \tag{2-11}$$

Spin due to the rigid body rotation, $\tilde{\boldsymbol{\Omega}}^e$, can then be defined and substituted in (2-5) as:

$$\tilde{\boldsymbol{L}}^* = \tilde{\boldsymbol{\Omega}}^e + \boldsymbol{R}^e \bar{\boldsymbol{L}}^p \boldsymbol{R}^{e^T}; \qquad \tilde{\boldsymbol{\Omega}}^e = \dot{\boldsymbol{R}}^e \boldsymbol{R}^{e^T} \tag{2-12}$$

The value of $\bar{\boldsymbol{L}}^p$ from (2-10) can then be substituted into (2-12) to obtain:

$$\tilde{\boldsymbol{L}}^* = \tilde{\boldsymbol{\Omega}}^e + \boldsymbol{R}^e \sum_{\chi=1}^{\mathrm{N}} \dot{\gamma}^\chi\, \bar{\boldsymbol{s}}^\chi \otimes \bar{\boldsymbol{m}}^\chi \boldsymbol{R}^{e^T} + \boldsymbol{R}^e \frac{1}{3} A_n \dot{\xi} \mathbf{1} \boldsymbol{R}^{e^T} \tag{2-13}$$

Again, since the void growth part of the velocity gradient is volumetric, rotation, $\boldsymbol{R}^e$, will not affect it. But the rotation of $\bar{\boldsymbol{s}}^\chi$ and $\bar{\boldsymbol{m}}^\chi$ by $\boldsymbol{R}^e$ will yield:

$$\tilde{\boldsymbol{L}}^* = \tilde{\boldsymbol{\Omega}}^e + \sum_{\chi=1}^{\mathrm{N}} \dot{\gamma}^\chi\, \tilde{\boldsymbol{s}}^\chi \otimes \tilde{\boldsymbol{m}}^\chi + \frac{1}{3} A_n \dot{\xi} \mathbf{1} \tag{2-14}$$

Here, $\tilde{\boldsymbol{s}}^\chi$ and $\tilde{\boldsymbol{m}}^\chi$ are given by:

$$\tilde{\boldsymbol{s}}^\chi = \boldsymbol{R}^e \bar{\boldsymbol{s}}^\chi \boldsymbol{R}^{e^T}, \qquad \tilde{\boldsymbol{m}}^\chi = \boldsymbol{R}^e \bar{\boldsymbol{m}}^\chi \boldsymbol{R}^{e^T} \tag{2-15}$$

The total velocity gradient can be decomposed into symmetric, $\boldsymbol{d}$ and skew, $\boldsymbol{w}$ parts given by:

$$\boldsymbol{l} = \boldsymbol{d} + \boldsymbol{w} \tag{2-16}$$

The rate of deformation tensor, $\boldsymbol{d}$ and the spin tensor, $\boldsymbol{w}$ can then be defined using (2-4) as:

$$\boldsymbol{d} = \mathrm{sym}\big(\dot{\boldsymbol{V}}^e \boldsymbol{V}^{e^{-1}}\big) + \boldsymbol{V}^{e^{-T}} \tilde{\boldsymbol{D}}^* \boldsymbol{V}^{e^{-1}}, \qquad \boldsymbol{w} = \mathrm{skew}\big(\dot{\boldsymbol{V}}^e \boldsymbol{V}^{e^{-1}}\big) + \boldsymbol{V}^{e^{-T}} \tilde{\boldsymbol{W}}^* \boldsymbol{V}^{e^{-1}} \tag{2-17}$$

Here, $\tilde{\boldsymbol{D}}^*$ and $\tilde{\boldsymbol{W}}^*$ are given by:

$$\tilde{\boldsymbol{D}}^* = \mathrm{sym}\big(\tilde{\mathbf{C}}^e \tilde{\boldsymbol{\Omega}}^e\big) + \boldsymbol{R}^e \bar{\boldsymbol{D}}^p \boldsymbol{R}^{e^T}, \qquad \tilde{\boldsymbol{W}}^* = \mathrm{skew}\big(\tilde{\mathbf{C}}^e \tilde{\boldsymbol{\Omega}}^e\big) + \boldsymbol{R}^e \bar{\boldsymbol{W}}^p \boldsymbol{R}^{e^T} \tag{2-18}$$

where $\tilde{\mathbf{C}}^e = \boldsymbol{V}^{e^T} \boldsymbol{V}^e$ and, in the next equation, $\bar{\boldsymbol{C}}^e = \boldsymbol{F}^{e^T} \boldsymbol{F}^e$ are the elastic right Cauchy-Green tensors, and $\bar{\boldsymbol{D}}^p$ and $\bar{\boldsymbol{W}}^p$ are defined below:

$$\bar{\boldsymbol{D}}^p = \mathrm{sym}(\bar{\boldsymbol{C}}^e \bar{\boldsymbol{L}}^p) = \mathrm{sym}\{\bar{\boldsymbol{C}}^e (\bar{\boldsymbol{L}}_s^p + \hat{\boldsymbol{L}}_v^p)\}, \qquad \bar{\boldsymbol{W}}^p = \mathrm{skew}\big(\bar{\boldsymbol{C}}^e \bar{\boldsymbol{L}}_s^p\big) \tag{2-19}$$

It can be seen in (2-19) that the rate of deformation tensor due to plastic deformation, $\bar{\boldsymbol{D}}^p$, depends on both the slip and void growth based plastic deformations, but the spin tensor only depends on the deformation due to slip, because void growth does not change the shape of the lattice.

The definition of $\tilde{\boldsymbol{L}}$ given in (2-11) along with $\tilde{\boldsymbol{D}} = \mathrm{sym}(\tilde{\boldsymbol{C}}^e \tilde{\boldsymbol{L}})$ can be used to get $\tilde{\boldsymbol{D}} = \boldsymbol{V}^{e^T} \boldsymbol{d} \boldsymbol{V}^e$ in which the value of $\boldsymbol{d}$ can then be substituted from (2-17) to get:

$$\tilde{\boldsymbol{D}} = \boldsymbol{V}^{e^T} \big(\mathrm{sym}\big(\dot{\boldsymbol{V}}^e \boldsymbol{V}^{e^{-1}}\big) + \boldsymbol{V}^{e^{-T}} \tilde{\boldsymbol{D}}^* \boldsymbol{V}^{e^{-1}}\big) \boldsymbol{V}^e \tag{2-20}$$

This relation can then be reduced with the help of $\dot{\widetilde{\boldsymbol{E}}}^e = \boldsymbol{V}^{e^T}\left(\dot{\boldsymbol{V}}^e\boldsymbol{V}^{e^{-1}}\right)\boldsymbol{V}^e$ obtaining the following relation:

$$\widetilde{\boldsymbol{D}} = \dot{\widetilde{\boldsymbol{E}}}^e + \widetilde{\boldsymbol{D}}^* = \overset{\nabla}{\widetilde{\boldsymbol{E}}}^e + \boldsymbol{R}^e\overline{\boldsymbol{D}}^p\boldsymbol{R}^{e^T} \tag{2-21}$$

Here, $\widetilde{\boldsymbol{D}}^*$ is then transformed into $\overline{\boldsymbol{D}}^p$ using $\boldsymbol{R}^e$ and $\overset{\nabla}{\widetilde{\boldsymbol{E}}}^e = \dot{\widetilde{\boldsymbol{E}}}^e + \widetilde{\boldsymbol{E}}^e\widetilde{\boldsymbol{\Omega}}^e - \widetilde{\boldsymbol{\Omega}}^e\widetilde{\boldsymbol{E}}^e$ is the Green-McInnis-Naghdi type rate of $\widetilde{\boldsymbol{E}}^e$ based on the elastic spin of lattice, $\widetilde{\boldsymbol{\Omega}}^e$.

In a similar way, the spin tensor in the intermediate configuration can be evaluated using the total spin tensor, $\boldsymbol{w}$ and then additively split into the elastic and the plastic part.

$$\widetilde{\boldsymbol{W}} = \boldsymbol{V}^{e^T}\boldsymbol{w}\boldsymbol{V}^e = \mathrm{skew}\left(\boldsymbol{V}^{e^T}\dot{\boldsymbol{V}}^e\right) + \widetilde{\boldsymbol{W}}^* \tag{2-22}$$

Anisotropic elasticity is used to calculate the 2nd Piola-Kirchhoff stress from the elastic strain tensor. Here, $\tilde{\mathbb{C}}^e$ is fourth-order stiffness tensor.

$$\widetilde{\boldsymbol{S}} = \tilde{\mathbb{C}}^e : \widetilde{\boldsymbol{E}}^e \tag{2-23}$$

The final relations for the plastic part of the deformation rate, containing both slip based and void growth parts, along with lattice rotation, can then be found using (2-10), (2-18) and (2-19). Here, $\widetilde{\boldsymbol{Z}}^\chi$ is the Schmid tensor in the intermediate configuration and is equal to $\tilde{\boldsymbol{s}}^\chi \otimes \widetilde{\boldsymbol{m}}^\chi$.

$$\widetilde{\boldsymbol{D}}^* = \mathrm{sym}\left(\widetilde{\mathbf{C}}^e\widetilde{\boldsymbol{\Omega}}^e\right) + \sum_{\chi=1}^{\mathrm{N}} \dot{\gamma}^\chi \mathrm{sym}\left(\widetilde{\boldsymbol{C}}^e\widetilde{\boldsymbol{Z}}^\chi\right) + \frac{1}{3}A_n\dot{\xi}\,\mathrm{sym}\left(\widetilde{\boldsymbol{C}}^e\right),$$
$$\widetilde{\boldsymbol{W}}^* = \mathrm{skew}\left(\widetilde{\mathbf{C}}^e\widetilde{\boldsymbol{\Omega}}^e\right) + \sum_{\chi=1}^{\mathrm{N}} \dot{\gamma}^\chi \;\mathrm{skew}\left(\widetilde{\boldsymbol{C}}^e\widetilde{\boldsymbol{Z}}^\chi\right) \tag{2-24}$$

An assumption of small elastic strains is made because their magnitude will be negligible compared to strains associated with plastic flow. The small elastic strains assumption for $\boldsymbol{V}^e$ was introduced using:

$$\boldsymbol{V}^e = \mathbf{1} + \boldsymbol{\epsilon}^e, \qquad \|\boldsymbol{\epsilon}^e\| \ll \mathbf{1} \tag{2-25}$$

The following results can be obtained using the above relations:

$$\dot{\boldsymbol{V}}^e = \dot{\boldsymbol{\epsilon}}^e, \qquad \boldsymbol{V}^{e^{-1}} = \mathbf{1} - \boldsymbol{\epsilon}^e + \mathcal{O}\|\boldsymbol{\epsilon}^e\|^2 \tag{2-26}$$

An infinitesimal difference will be left between the final state and the unloaded intermediate state. The following relations can be derived using these approximations along with the consideration that higher-order terms, $\mathcal{O}\|\boldsymbol{\epsilon}^e\|^2$ and terms like $(\blacksquare)\boldsymbol{\epsilon}^e$ and $\boldsymbol{\epsilon}^e(\blacksquare)$ will be negligible:

$$\widetilde{\boldsymbol{D}} \approx \boldsymbol{d}, \qquad \widetilde{\boldsymbol{W}} \approx \boldsymbol{w}, \quad \mathrm{skew}\left(\boldsymbol{V}^{e^T}\dot{\boldsymbol{V}}^e\right) \approx \mathrm{skew}(\dot{\boldsymbol{\epsilon}}^e\boldsymbol{\epsilon}^e)$$
$$\widetilde{\boldsymbol{C}}^e \approx \mathbf{1}, \qquad \widetilde{\boldsymbol{E}}^e \approx \boldsymbol{\epsilon}^e, \qquad \widetilde{\boldsymbol{S}} \approx \boldsymbol{\tau}$$
$$\widetilde{\boldsymbol{D}}^* \approx 2\mathrm{sym}\left(\boldsymbol{\epsilon}^e\widetilde{\boldsymbol{\Omega}}^e\right) + \sum_{\chi=1}^{\mathrm{N}} \dot{\gamma}^\chi \mathrm{sym}\left(\widetilde{\boldsymbol{Z}}^\chi\right) + \frac{1}{3}A_n\dot{\xi}\mathbf{1},$$
$$\widetilde{\boldsymbol{W}}^* \approx \widetilde{\boldsymbol{\Omega}}^e + \sum_{\chi=1}^{\mathrm{N}} \dot{\gamma}^\chi \;\mathrm{skew}\left(\widetilde{\boldsymbol{Z}}^\chi\right) \tag{2-27}$$

After incorporating the results given in (2-27), (2-21) and (2-22) can then be written as:

$$\boldsymbol{d} = \overset{\nabla}{\boldsymbol{\epsilon}}^e + \widetilde{\boldsymbol{D}}^p, \qquad \overset{\nabla}{\boldsymbol{\epsilon}}^e = \dot{\boldsymbol{\epsilon}}^e + \boldsymbol{\epsilon}^e\widetilde{\boldsymbol{\Omega}}^e - \widetilde{\boldsymbol{\Omega}}^e\boldsymbol{\epsilon}^e$$
$$\boldsymbol{w} = -\mathrm{skew}(\dot{\boldsymbol{\epsilon}}^e\boldsymbol{\epsilon}^e) + \widetilde{\boldsymbol{\Omega}}^e + \widetilde{\boldsymbol{W}}^p, \qquad \widetilde{\boldsymbol{\Omega}}^e = \dot{\boldsymbol{R}}^e\boldsymbol{R}^{e^T} \tag{2-28}$$

The elastic constitutive equation given in (2-23) can then be updated to:

$$\boldsymbol{\tau} = \tilde{\mathbb{C}}^e : \boldsymbol{\epsilon}^e \tag{2-29}$$

And the plasticity relations can then be written as:

$$\widetilde{\boldsymbol{D}}^p = \boldsymbol{R}^e \overline{\boldsymbol{D}}^p \boldsymbol{R}^{e^T} = \sum_{\chi=1}^{N} \dot{\gamma}^{\chi} \mathrm{sym}\big(\widetilde{\boldsymbol{Z}}^{\chi}\big) + \frac{1}{3} A_n \dot{\xi} \mathbf{1}$$

$$\widetilde{\boldsymbol{W}}^p = \boldsymbol{R}^e \overline{\boldsymbol{W}}^p \boldsymbol{R}^{e^T} = \sum_{\chi=1}^{N} \dot{\gamma}^{\chi} \mathrm{skew}\big(\widetilde{\boldsymbol{Z}}^{\chi}\big) \tag{2-30}$$

For the case of elastically isotropic and rigid metals, the above porous crystal plasticity formulation converges to a simpler version which is given in Appendix A.

## 2.2 Void volume fraction evolution

### 2.2.1 Nucleation and growth

The void volume fraction $f$, normalised by the initial value of void volume fraction $f_0$, is treated as a strain like quantity, $\xi$ that represents the volumetric plastic deformation of a material point due to void growth. The void volume fraction evolution is treated in two steps, namely nucleation/growth followed by coalescence. The nucleation and growth stage, $\xi_g$ is found to be a function of the applied equivalent strain, $\epsilon_{eq}$, the stress triaxiality, $X$, and the Lode parameter, $L$ among others. This is based on the experimental findings and the RVE studies discussed in the introduction. Power law functions are used for these quantities to relate their effect on the evolution of $\xi_g$, given by:

$$\dot{\xi}_g = \frac{\dot{f}}{f_0} = \frac{(1+X)^{\mathcal{A}} \left(\frac{\epsilon_{eq}}{\mathcal{C}}\right)^{\mathcal{B}}}{(1+L)^{\mathcal{D}}} \left(\mathcal{A}\frac{\dot{X}}{1+X} + \mathcal{B}\frac{\dot{\epsilon}_{eq}}{\epsilon_{eq}} - \mathcal{D}\frac{\dot{L}}{1+L}\right), \qquad \mathcal{D} = \begin{cases} \mathcal{D}, & L > 0 \\ 0, & L \le 0 \end{cases} \tag{2-31}$$

The condition in (2-31) implies that the effect of the Lode parameter will only be present when its value is positive and greater than 0, otherwise it will not affect void growth. This feature is based on the findings of representative volume element studies that showed that void growth slows down at values of $L$ higher than 0 in contrast to void growth when $L$ is negative [41], [52], [53], [55].

Stress triaxiality is defined as the ratio between the hydrostatic stress, $p_\tau$, and von Mises stress and is given by $X = \frac{p_\tau}{\sqrt{\frac{3}{2}\mathrm{dev}\boldsymbol{\tau}:\mathrm{dev}\boldsymbol{\tau}}}$, equivalent strain is defined as $\epsilon_{eq} = \sqrt{\frac{2}{3}\mathrm{dev}\boldsymbol{\epsilon}:\mathrm{dev}\boldsymbol{\epsilon}}$ and the Lode parameter is given by $L = -\frac{27}{2}\frac{|\mathrm{dev}\boldsymbol{\tau}|}{\sqrt{\frac{3}{2}\mathrm{dev}\boldsymbol{\tau}:\mathrm{dev}\boldsymbol{\tau}}}$.

The values of material parameters $\mathcal{A}$ and $\mathcal{C}$, in (2-31), are found to be the functions of the phase boundary inclination angle $(pbi)$ in the case of void growth at the interface of the two phases in dual phase alloys [2]. They are related to $pbi$ using:

$$\mathcal{A} = \mathcal{E}\mathrm{sech}(\mathcal{F}(pbi) - \mathcal{G})$$

$$\mathcal{C} = \mathcal{H}\mathrm{sech}(\mathcal{I}(pbi) - \mathcal{J}) \tag{2-32}$$

Here, $\mathcal{E}$, $\mathcal{F}$, $\mathcal{G}$, $\mathcal{H}$, $\mathcal{I}$ and $\mathcal{J}$ are material parameters. Their values are found using RVE calculations [1], [2] and depend on the mechanical behaviour of the two phases and the value of the initial porosity. Here, it should be noted that, for the case of single crystals, the values of $\mathcal{A}$ and $\mathcal{C}$ will be constant and not functions of $pbi$.

### 2.2.2 Coalescence

The void growth rate during the coalescence stage is accelerated using two parameters, $a_1$ and $a_2$ that control the rate of void growth and the change in the rate of void growth respectively. A similar formulation has been used by researchers in the past [29], [39], [50]. It results in a relation which is a function of $\xi_g$, and its cut-off value, $\xi_{gc}$ at which the mechanism will shift from void growth to coalescence, which is given below:

$$\xi_{coal} = \xi_{gc} + a_1\left(\xi_g^{a_2} - \xi_{gc}^{a_2}\right) \tag{2-33}$$

The condition for the choice between $\xi_g$ and $\xi_{coal}$ is based on the accumulated value of $\xi$; if the value is less than $\xi_{gc}$ then it will be calculated using the void nucleation and growth relation given in(2-31), otherwise, (2-33) will be used, and therefore:

$$\xi = \begin{cases} \xi_g, & \xi < \xi_{gc} \\ \xi_{coal}, & \xi \geq \xi_{gc} \end{cases} \tag{2-34}$$

## 2.3 Updated flow rule

The flow rule used in the formulation of a non-porous crystal plasticity theory by Marin [56] was updated to incorporate the effects of void growth in porous single crystals. The lattice tends to soften with the evolution of the void volume fraction because the presence of a void induces a stress concentration around the void. This softening is added in the existing flow rule as an exponential function of $\xi$, which will reduce the slip system strength as the void volume fraction increases. Here, the coefficient $\mathscr{s}_1$ is a material parameter (a similar relation has been used previously [29]) which regulates this effect of softening. The resulting relation is given as:

$$\dot{\gamma}^{\chi} = \dot{\gamma}_0 \left[\frac{|\tau^{\chi}|}{\kappa_s^{\chi} \exp(-\mathscr{s}_1 \xi)}\right]^{1/m} sign(\tau^{\chi}) \tag{2-35}$$

## 2.4 Updated hardening evolution

The hardening evolution law, given in (2-37), is the same as the one used by Marin [56], but the value of the critical resolved shear stress (CRSS), $\kappa_0^{\chi}$ was updated to be a function of stress triaxiality and PBI. The equivalent stress – equivalent strain results obtained from the RVE simulations [2] showed that yielding starts at a lower stress magnitude in higher stress triaxialities as compared to the uniaxial case. An exponential function of stress triaxiality is used to model this behaviour (see (2-36)), with $\mathscr{s}_2$ as a material parameter which is used to scale this effect. This function is made to have no effect in the uniaxial case where $X$=1/3.

As discussed elsewhere [2], the mechanical behaviour of a bicrystal of $\alpha$ and $\beta$ phases of titanium alloys is significantly different than the homogenised response of the non-porous crystals of $\alpha$ and $\beta$ phases (i.e. the volume averaged stress, given in Appendix B.4). This is because of the fact that the deformation of one phase in the bicrystal is affected by the other phase. Also, the effect of PBI further complicates the situation as the response of the bicrystal will then depend on the relative stiffness of the single crystals of the $\alpha$ and $\beta$ phases. The value of PBI defines whether the two crystals have their deformations (1) constrained to be equal to the total deformation ($pbi$ = 90°), (2) are such that the sum of their deformations in the major loading direction is equal to the total deformation ($pbi$ = 0°), or (3) such that the behaviour is in between (0° < $pbi$ < 90°). This anisotropy affects void growth and the equivalent stress – equivalent strain response in a way that cannot be captured by existing crystal plasticity formulations. In order to capture this effect, a simple linear function of $pbi$ (taken in radians) is used. A material parameter, $\mathscr{s}_3$ normalised by the CRSS of a selected slip system, $\kappa_{0|ref}^{\chi}$, is used to scale this effect. The value of $\kappa_0^{\chi}$ of a prismatic slip system of the $\alpha$ phase is used as $\kappa_{0|ref}^{\chi}$. The value of $\mathscr{s}_3$ will be zero for the case of single crystals.

The above results in a relation that takes into account these effects and is:

$$\kappa_0^{\chi} = \left\{\kappa_0^{\chi}\left(1 + \frac{\mathscr{s}_3(pbi)}{\kappa_{0|ref}^{\chi}}\right)\right\}\exp\left(-\mathscr{s}_2\left|X - \frac{1}{3}\right|\right) \tag{2-36}$$

The evolution law for the hardening of a slip system [56] is given by:

$$\dot{\kappa}_s^{\chi} = h_0\left(\frac{\kappa_{s,S} - \kappa_s}{\kappa_{s,S} - \kappa_{s,0}}\right)\sum_{\chi=1}^{N}|\dot{\gamma}^{\chi}|, \qquad \kappa_{s,S} = \kappa_{s,S0}\left[\frac{\sum_{\chi}|\dot{\gamma}^{\chi}|}{\dot{\gamma}_{s0}^{\chi}}\right]^{\frac{1}{m'}} \tag{2-37}$$

where $h_0$, $\kappa_{s,0}$, $\kappa_{s,S0}$, $\dot{\gamma}_{s0}^{\chi}$ and $m'$ are material parameters. The value of $\kappa_0^{\chi}$ will be used during integration of (2.37).

Details of numerical implementation are given in Appendix B.

# 3 Parametric assessment of the model

An extensive parametric assessment is carried out to study the capability of the model. Elastic and plastic parameters including, those required for the original flow rule, the hardening law, and the volume fractions of $\alpha$ and $\beta$ phases of the titanium alloy (Ti-10V-2Fe-3Al) are required for this study. The values of these parameters are extracted from the RVE study [2]. The values of parameters required in (2-31) and (2-32) have already been found by calibrating the void growth model in (2-31) (with $D$=0) using the RVE results [1]. These two sets of parameters are given in Table 3-1 and Table 3-2. The value of $\mathcal{D}$ is set to 0 because comparison will be made with the RVE study results for the Lode parameter value of -1.

*Table 3-1: Material parameters for CPFEM of Ti-1023 α-β phases*

| **$\alpha$ phase Properties** | | | | | | | | |
|---|---|---|---|---|---|---|---|---|
| Elastic Properties | $C_{11}$ | $C_{12}$ | $C_{13}$ | $C_{33}$ | $C_{44}$ | | | |
| (GPa) | 143.0 | 94.0 | 49.3 | 191.0 | 18.0 | | | |
| Plastic Properties | $\dot{\gamma}_0$ | $m$ | $h_0$ | $\kappa_0^{\chi}$ | $\kappa_{s,0}$ | $\kappa_{s,S0}$ | $\dot{\gamma}_{S0}$ | $m'$ |
| | | | | (MPa) | | | | |
| Basal | 0.01 | 0.05 | 10 | 190 | 1 | 100 | $5\times10^{10}$ | 0.005 |
| Prismatic | 0.01 | 0.05 | 10 | 160 | 1 | 60 | $5\times10^{10}$ | 0.005 |
| Pyramidal | 0.01 | 0.05 | 10 | 400 | 1 | 420 | $5\times10^{10}$ | 0.005 |
| **$\beta$ phase Properties** | | | | | | | | |
| Elastic Properties | $C_{11}$ | $C_{12}$ | $C_{44}$ | | | | | |
| (GPa) | 120.0 | 108.0 | 30.0 | | | | | |
| Plastic Properties | $\dot{\gamma}_0$ | $m$ | $h_0$ | $\kappa_0^{\chi}$ | $\kappa_{s,0}$ | $\kappa_{s,S0}$ | $\dot{\gamma}_{S0}$ | $m'$ |
| | | | | (MPa) | | | | |
| $\{110\}\langle111\rangle$ | 0.1 | 0.05 | 10 | 150 | 1 | 50 | $5\times10^{10}$ | 0.005 |
| $\{110\}\langle112\rangle$ | 0.1 | 0.05 | 10 | 170 | 1 | 75 | $5\times10^{10}$ | 0.005 |
| $\{110\}\langle123\rangle$ | 0.1 | 0.05 | 10 | 200 | 1 | 120 | $5\times10^{10}$ | 0.005 |

*Table 3-2: Parameters of the model calibrated for $f_0$=0.01, Ti-1023 alloy*

| $\mathcal{B}$ | $\mathcal{D}$ | $\mathcal{E}$ | $\mathcal{F}$ | $\mathcal{G}$ | $\mathcal{H}$ | $\mathcal{I}$ | $\mathcal{J}$ |
|---|---|---|---|---|---|---|---|
| 1.20 | 0 | 5.30 | 1.20 | 1.25 | 7.00 | 1.80 | 2.50 |

The detailed procedure followed to obtain the values of parameters given in Table 3-1 and Table 3-2 are found elsewhere [1], [2], [57], and not repeated here for brevity. The method for the identification of the rest of the parameters, i.e. for the plastic deformation due to void growth and its effect on slip-based plasticity, are presented next. The parameters $A_n$, $\mathscr{s}_1$, $\mathscr{s}_2$, $\mathscr{s}_3$, $\xi_{gc}$, $a_1$ and $a_2$, used in (2-8) and (2-33) to (2-36) were tested for their effects on the equivalent stress – equivalent strain response of a bicrystal with a void, and on the evolution of normalised void volume fraction with applied equivalent strain. The effect of all these parameters is studied at two levels of applied stress triaxiality, $X$= 1/3 and 3, and at $pbi$=90°, except for the parameter $\mathscr{s}_2$ whose results are presented at $X$= 1 and 3. The effect of the parameter $\mathscr{s}_3$ is investigated at $pbi$=30° and 90°. Performance of the model for different values of stress triaxialities at two different values of PBIs, and for different values of PBIs at two different values of stress triaxialities is evaluated in terms of equivalent stress – equivalent strain response and normalised void volume fraction evolution (Figure 3-3 and Figure 3-4). The values of the parameters at which this study was carried out and the values of stress triaxialities and PBIs at which the performance was assessed are given in Table 3-3.

*Table 3-3: Parameter and their values chosen for parametric study (* value of the parameter which is used while testing another parameter)*

| | | | | | | | Tested at | |
|---|---|---|---|---|---|---|---|---|
| Parameter | | | | | | | $X$ | $pbi$ |
| $A_n$ | 0.0 | 0.01 * | 0.02 | 0.03 | 0.05 | 0.1 | 1/3, 3 | 90° |
| $\mathscr{s}_1$ | 0.0 | 0.01 * | 0.02 | 0.03 | 0.05 | 0.1 | 1/3, 3 | 90° |
| $\mathscr{s}_2$ | 0.0 | 0.1 * | 0.2 | 0.3 | 0.5 | 1.0 | 1, 3 | 90° |
| $\mathscr{s}_3$ (MPa) | 0 | 10 | 20 | 50 | | | 1/3, 3 | 30°, 90° |
| $\xi_{gc}$ | 1.1 | 1.11 | 1.12 | 1.13 | 1.15 * | 1.2 | 1/3 | 90° |
| | 7.50 * | 7.75 | 8.00 | 8.50 | 9.00 | 10.0 | 3 | |
| $a_1$ | 1 | 10 | 20 | 30 | 50 | 100 * | 1/3 | 90° |
| | 1.0 | 2.0 * | 3.0 | 5.0 | 10.0 | | 3 | |
| $a_2$ | 1.0 | 1.1 | 1.2 | 1.3 | 1.5 | 2.0 | 1/3, 3 | 90° |
| $X$ | 1/3 | 1/2 | 3/4 | 1 | 2 | 3 | | 30°, 90° |
| $pbi$ | 0° | 30° | 45° | 60° | 75° | 90° | 1/3, 3 | |

It was found that the effect of a certain parameter may differ considerably at the two values of stress triaxiality. For example, the effect of $\mathscr{s}_1$ on the results is very prominent for the case of $X$=3, unlike at $X$=1/3. This is because of a very small difference between the results at different values of a given parameter (except $A_n$) or because of the physics involved at these different stress triaxialities (in the case of $A_n$). The effect of individual parameters is discussed in the following.

Figure 3-1 (a) and (b) shows the effect of the parameter $A_n$ at $X$=1/3 and 3. It can be inferred that this parameter does not have a significant effect for either of these two stress triaxiality values, but it is almost indiscernible in the case of $X$=1/3. $A_n$ is the scaling factor for the strain like parameter characterising void growth $\xi$, which in turn affects the volumetric part of the stress as per the relation given in (2-8). Since $\xi$ in the case of $X$=1/3 is very small (Figure 3-1 (a)), it does not have a strong effect on the equivalent stress – equivalent strain response. In contrast, a significant effect of $A_n$ can be observed in the case of $X$=3 (Figure 3-1 (b)). The effect of $A_n$ on void growth can also be observed at both stress triaxialities in Figure 3-1 (a) and (b), and is found to be insignificant, which is a desirable condition. Figure 3-1 (c) and (d) show the effect of the parameter $\mathscr{s}_1$ at $X$=1/3 and 3. This parameter helps to simulate the effect of stress concentrations around the void which results in the softening of the crystal, by scaling the CRSS of the crystal. This parameter is a multiplier in the exponential function of $\xi$ given in (2-35) which controls the rate of softening with void growth. In the case of $X$=1/3, the magnitude of $\xi$ is small (Figure 3-1 (c)) and that is why the amount of softening is small. However, for the same values of $\mathscr{s}_1$ in $X$=3 case (Figure 3-1 (d)), softening is very high which is due to the higher magnitude of $\xi$. This parameter does not have a significant effect on void growth as can be seen in Figure 3-1 (c) and (d), which is a desirable situation.

The effect of the parameter $s_2$ on the equivalent stress – equivalent strain response and void growth is shown in Figure 3-1 (e) and (f) at $X$=1 and 3, respectively. This parameter is a multiplier in an exponential function of stress triaxiality which scales the CRSS of a crystal with the change in stress triaxiality as per the relation given in (2-36). This relation is formulated such that there is no effect in the case of $X$=1/3 but the value of CRSS decreases exponentially with the increase in stress triaxiality and $s_2$ scales the decrease. The reduction in the value of CRSS is smaller for $X$=1, but a large reduction was observed for $X$=3 for the same values of $s_2$. The parameter $s_3$, having dimension of stress, was introduced to account for the change in strength of a bicrystal with a change in $pbi$ which was observed during the RVE study [2]. Its effects on equivalent stress – equivalent strain and void growth are shown in Figure 3-1 (g) and (h) for $pbi$=30° and 90°, respectively at $X$=1/3 and 3. The parameter $s_3$ is a multiplier of $pbi$ with which the value of CRSS is scaled linearly as $pbi$ increases from 0° to 90°, (refer to (2-36)). It can be inferred from Figure 3-1 (g) and (h) that the value of CRSS increased with an increase in the value of $s_3$ at both values of stress triaxiality, and this increase is higher in the case of $pbi$=90°. This parameter has no effect on void growth, which is a desirable situation.

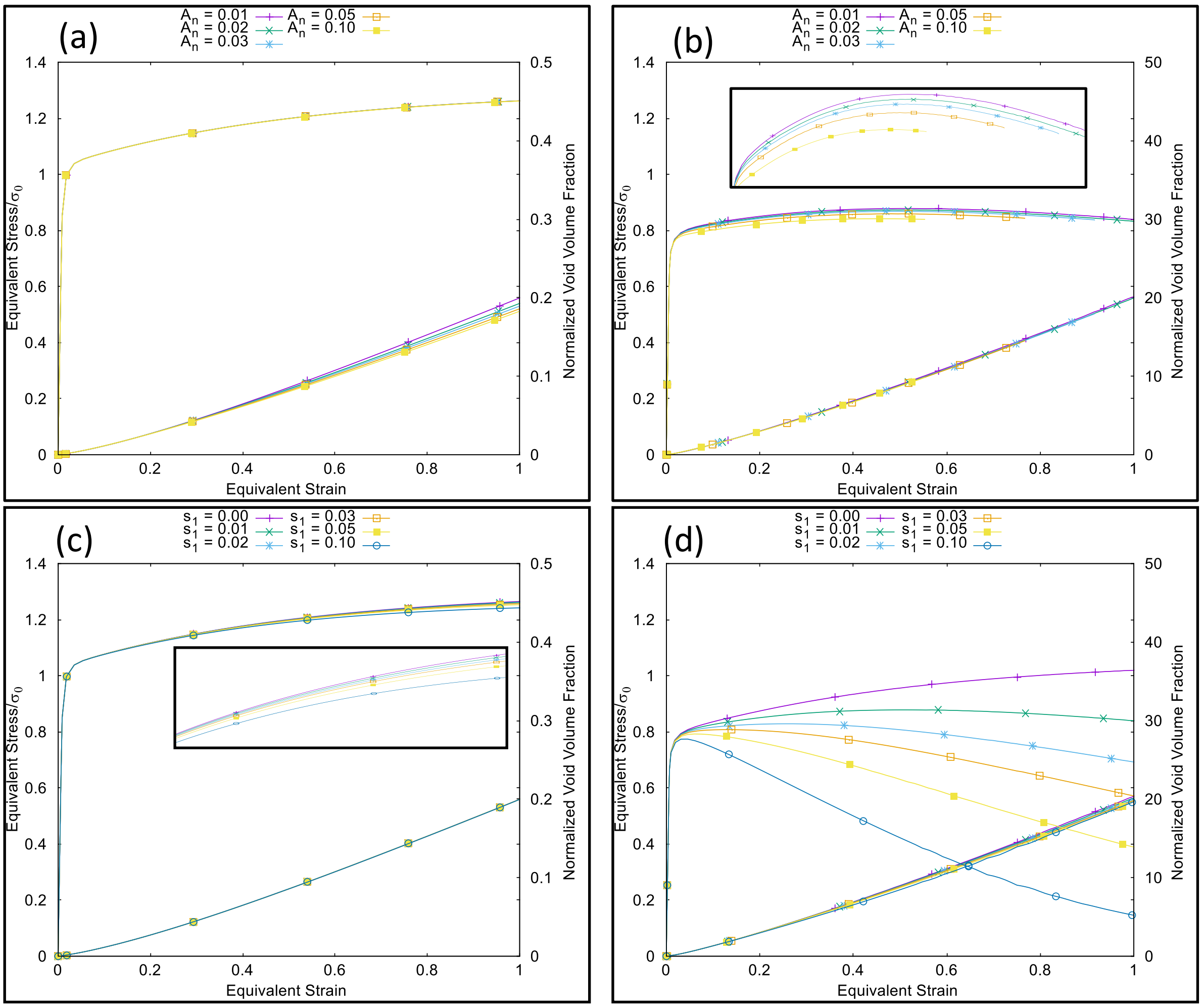

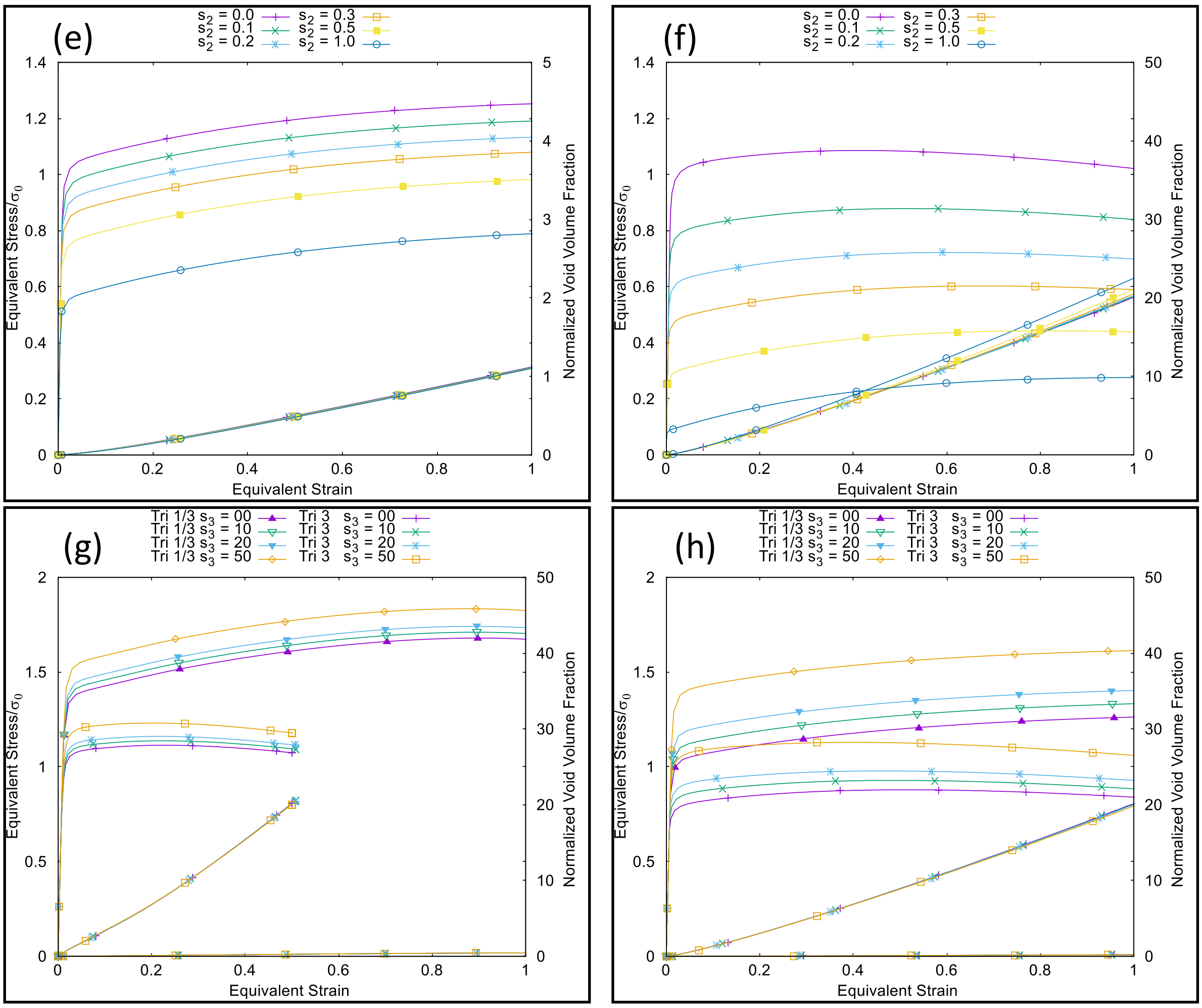

*Figure 3-1: Effect of parameter (a) $A_n$ at $X$=1/3 and (b) $X$=3, (c) $s_1$ at $X$=1/3 and (d) $X$=3, (e) $s_2$ at $X$=1 and (f) $X$=3, and (g) $s_3$ at $pbi$=30° and (h) $pbi$=90° on void growth and equivalent stress – equivalent strain response*

The effect of void coalescence parameters on equivalent stress – equivalent strain and evolution of normalised void volume fraction is shown in Figure 3-2 (a) to (f) at $X$=1/3 and 3. The parameter $\xi_{gc}$ is the value of $\xi$ at which the mechanism of void volume fraction evolution shifts from simple growth to coalescence. Void coalescence is modelled as an accelerated void growth in this formulation. This can be observed in Figure 3-2 (a) and (b) where there is a sudden increase in void growth at different values of $\xi$ as $\xi_{gc}$ was changed. This switch from void growth to void coalescence accelerated the softening at the corresponding value of equivalent strain. The softening is more pronounced in the case of $X$=3 because the magnitudes of $\xi$ are large and after the onset of coalescence, they become even larger. The function of parameters $a_1$ and $a_2$ is to accelerate the void growth to simulate void coalescence. Their effects on normalised void volume fraction evolution and equivalent stress – equivalent strain response are shown in Figure 3-2 (c-d) and (e-f) at $X$=1/3 and 3. Apparently, their effects seem to be similar but, in combination, they are used to control the rate of change of normalised void volume fraction evolution. The parameter $a_1$ simply scales the void coalescence rate, but careful use of suitable values of $a_2$ can control the shape of the void volume fraction evolution curve.

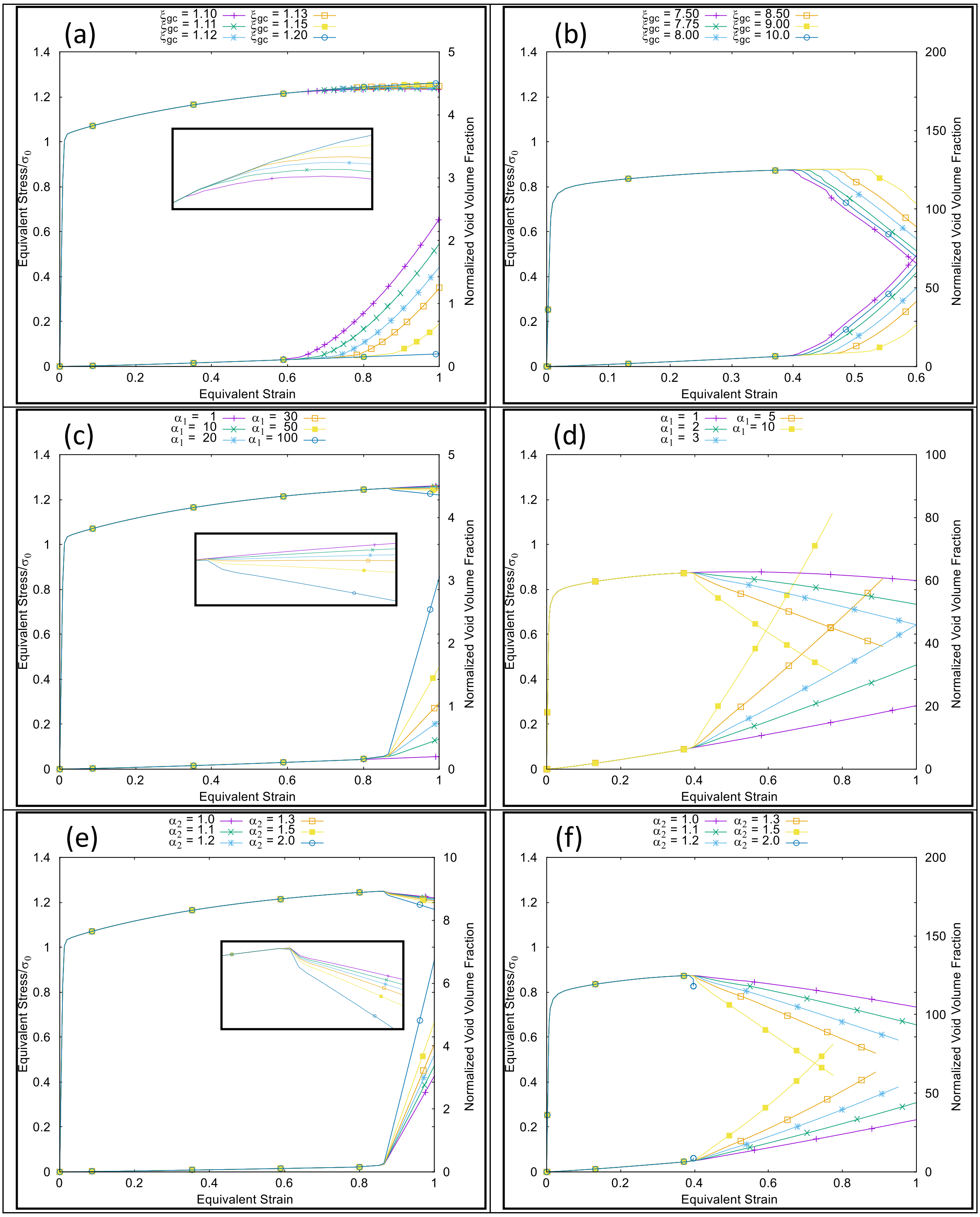

*Figure 3-2: Effect of parameter (a) $\xi_{gc}$ at X=1/3 and (b) X=3, (c) $a_1$ at X=1/3 and (d) X=3, and (e) $a_2$ at X=1/3 and (f) X=3 on void growth and equivalent stress – equivalent strain response*

The performance of the model at various values of stress triaxiality and for $pbi$=30° and 90° was investigated and the results are shown in Figure 3-3 (a) and (b), respectively. It can be seen that void growth has increased with the increase in stress triaxiality for both PBIs. This increase is higher at $pbi$=30° as compared to $pbi$=90°, in agreement with the results of the RVE study [2]. Also, it can be seen in the equivalent stress – equivalent strain response that the case with $pbi$=30° has higher yield stresses at all stress triaxialities than the $pbi$=90° case, because the strengths of the bicrystal are different in these two orientations.

Figure 3-4 (a) and (b) shows the performance of the model at different PBIs at $X$=1/3 and 3, respectively. It can be seen that the yield stresses are changing with variation in PBI; this is due to the fact that (i) CRSS depends on the crystal orientation, and in order to change the PBI, the bicrystal was rotated which changes their crystal orientation; and (ii) the CRSS was made a function of PBI to simulate the results obtained from RVE study [2]. The void growth in the cases of $X$=1/3 and 3 follows the trends seen in the RVE study [2].

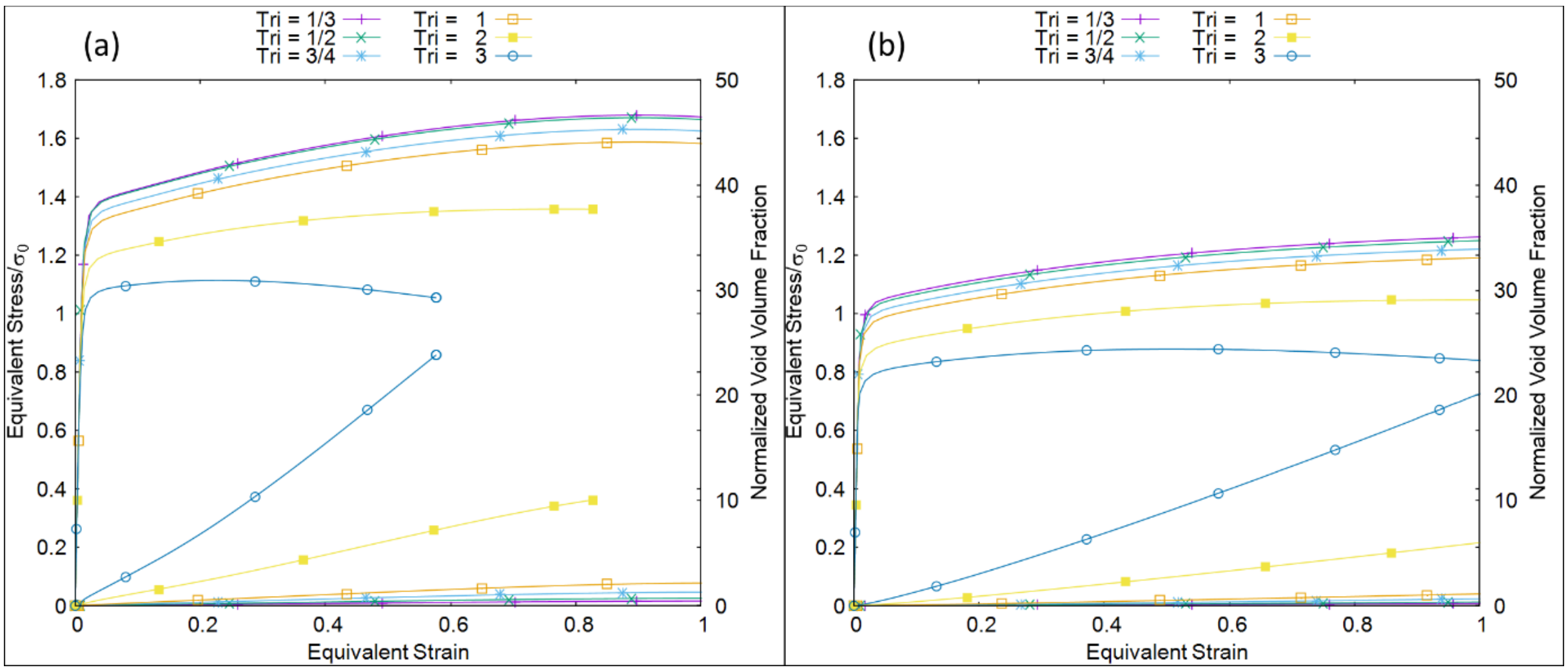

*Figure 3-3: Effect of applied stress triaxiality ($X$) on void growth and equivalent stress – equivalent strain response at (a) $PBI$=30° and (b) $PBI$=90°*

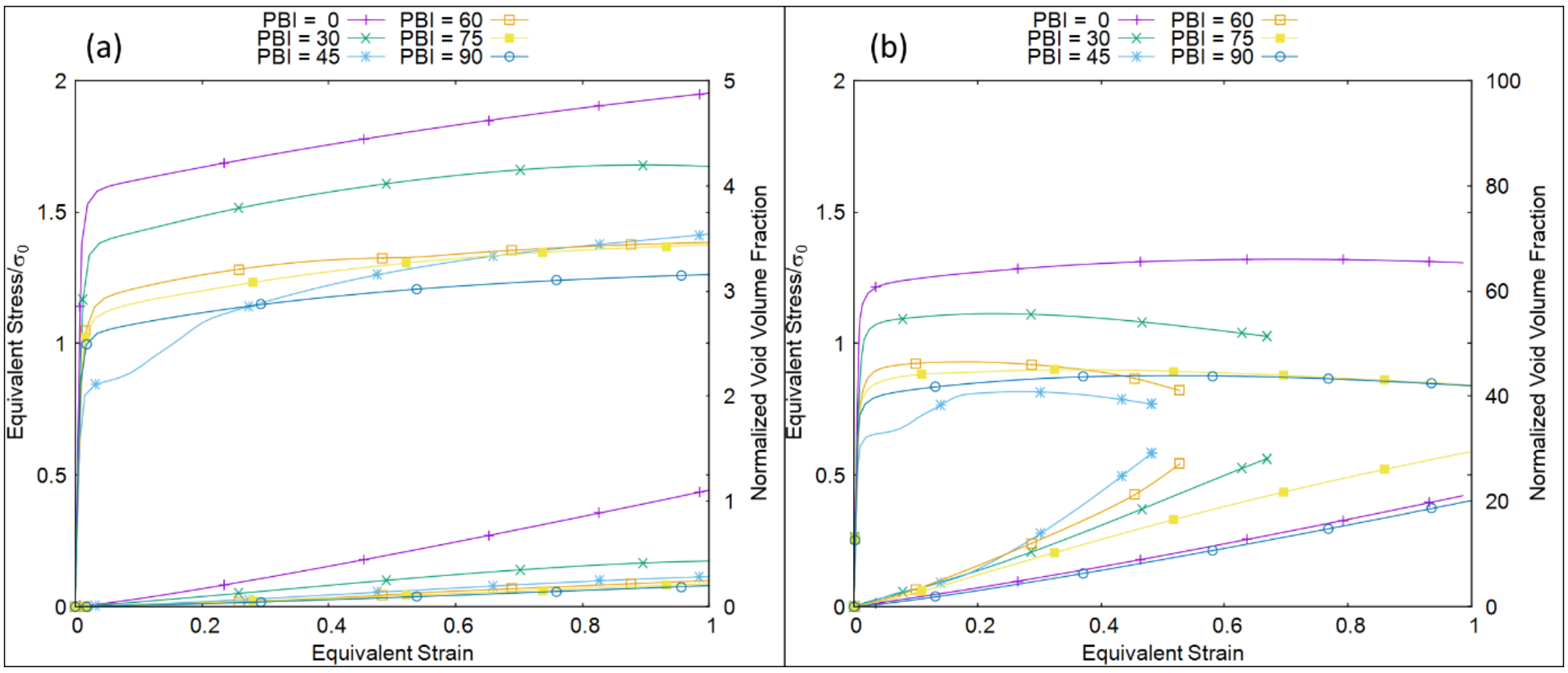

*Figure 3-4: Effect of phase boundary inclination (pbi) on void growth and equivalent stress – equivalent strain response at (a) $X$=1/3 and (b) $X$=3*

# 4 Model verification and validation

To demonstrate model performance, validation and verification, results from the RVE study of void growth in a bicrystal of a $\alpha$-$\beta$ titanium alloy (Ti-1023) [2] are compared with those for the model presented above. Sets of material parameters required for this study include those for non-porous single crystal plasticity, void nucleation and growth. The values of the first set of material parameters (for nonporous crystal plasticity) are kept the same as used for the RVE simulation of the bicrystal of $\alpha$-$\beta$ phases of Ti-1023 titanium alloy [1], [2] and are given in Table 3-1 and Table 3-2. The values of material parameters required for void nucleation, growth and coalescence were calibrated using the results of the RVE study [2] through an inverse modelling approach and are given in Table 4-1. The volume fraction of phases is kept at 0.5 for each of the phases. A comparison is made for 4 $pbi$ angles and the crystal orientations of each of the two phases of these 4 $pbi$ are given in Table 4-2.

*Table 4-1: Parameters for the porous plasticity model calibrated via the results of the RVE study of Ti-1023 alloy* [2]

| $A_n$ | $s_1$ | $s_2$ | $s_3$ (MPa) | $\xi_{gc}$ | $a_1$ | $a_2$ |
|---|---|---|---|---|---|---|
| 0.02 | 0.018 | 0.16 | 70 | 2.4 | 10 | 1.1 |

*Table 4-2: The Euler angles of $\alpha$ and $\beta$ phases for different phase boundary inclinations tested*

| No. | $pbi$ | $\alpha$-phase | | | $\beta$-phase | | |
|---|---|---|---|---|---|---|---|
| | | $\Psi$ | $\Theta$ | $\phi$ | $\Psi$ | $\Theta$ | $\phi$ |
| 1 | 90° | 0° | 0° | 180° | 324.74° | 45.00° | 180.00° |
| 2 | 60° | 180° | 30° | 0° | 289.73° | 25.70° | 138.27° |
| 3 | 30° | 180° | 60° | 0° | 231.59° | 31.40° | 073.67° |
| 4 | 0° | 180° | 90° | 0° | 210.00° | 54.74° | 045.00° |

The comparison is made on the basis of the equivalent stress – equivalent strain response for each of the 4 $pbi$ angles at $X$=1, 2 and 3, which are given in Figure 4-1 (a-d). It can be seen that the model predictions are in good agreement with the RVE response.

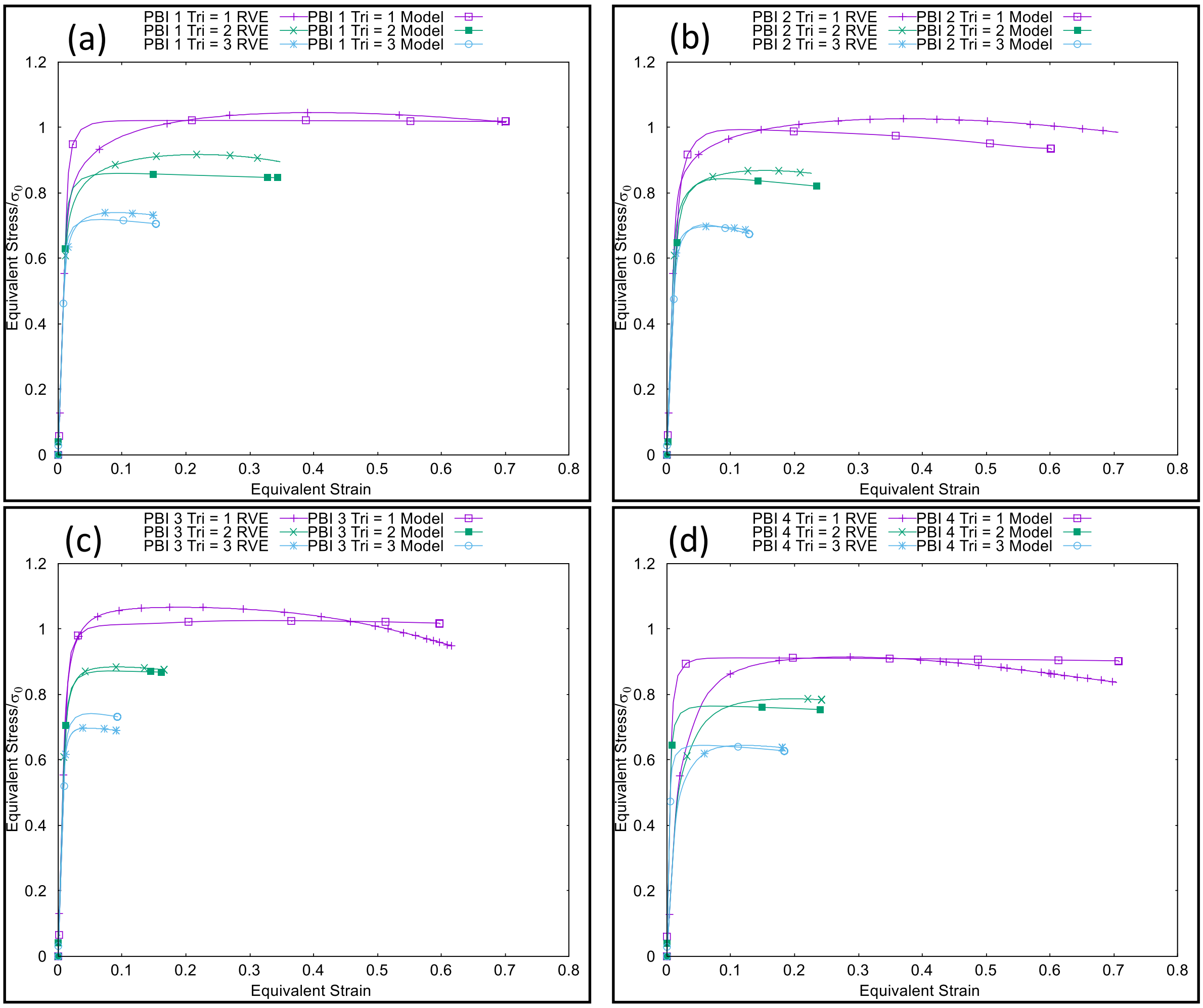

*Figure 4-1: A comparison of the equivalent stress – equivalent strain responses of the RVE study and the prediction of the developed model in (a) PBI 1, (b) PBI 2, (c) PBI 3, and (d) PBI 4*

# 5 Application of the constitutive model

To demonstrate model application and capability the constitutive model is assessed for a real example of sheet metal forming. In metal forming applications, the forming limit diagram/forming limit curve (FLC) is used for designing the process. Plasticity models are calibrated from data obtained from the FLC and then these models are used to develop and optimise the manufacturing process on a case-by-case basis.

Most of these models are at the macroscale and they are unable to provide insight into micromechanical aspects of the forming process. There has been a limited effort in the past to get insight into the micromechanics of ductile failure (for details see references [58], [59] and references therein). Viatkina *et al.* [58] used strain localisation as the criterion for failure in the polycrystalline aggregate and the response of each of the single crystals in the aggregate was solved using the CPFEM method. Gupta *et al.* [59] used the Marciniak-Kuczynski (M-K) model at the macroscopic scale of the polycrystalline aggregate to simulate sheet-necking along with the CPFEM formulation for the non-porous single crystals. The constitutive model presented here tries to overcome this shortcoming by extending the current crystal plasticity finite element method to incorporate the effects of crystal anisotropy and the phase boundary orientation in the context of porous crystal plasticity.

In order to show the model capability, the FLC results for a Ti-6Al-4V titanium alloy with a mill-annealed structure are used [60]. A small amount of $\beta$ phase (6.14%) was present in the microstructure within a large quantity of $\alpha$ phase (93.86%) [60]. It can be seen in the tensile stress – strain response, shown in Figure 5-1 that the material underwent considerable plastic deformation, and necking was also reported before failure [60]. Void nucleation, growth and coalescence was reported as the failure mechanism in this alloy [6]. It was also reported that voids nucleated on $\alpha-\beta$ phase boundaries [7]. Hence, our formulation of void growth on the interface of the $\alpha-\beta$ phases is applicable for this case. The values of material parameters for the void nucleation and growth model are assumed to be same as given in Table 3-2. For the rest of the parameters, an inverse modelling approach, based on published results [60], was used to calibrate our model.

## 5.1 Parameter identification for Ti-6Al-4V alloy

Uniaxial tensile test results [60] were used first to calibrate the material parameters required for the flow and hardening laws. This test was carried out at room temperature, at a test speed of 2mm/min, and for specimens cut at 0° to the rolling direction. It was reported that the microstructure had 94% by volume of $\alpha$ phase and 6% of $\beta$ phase.An inverse modelling approach [2] was used to identify the values of material parameters for use in the constitutive model presented in this work and are given in Table 5-1. A comparison between the stress-strain response from the experiment and that due to our constitutive model is given in Figure 5-1.

*Table 5-1: The material parameters for α-β phases of Ti-6Al-4V for the constitutive model*

| **$\alpha$ phase Properties, $v_1$=0.94** | | | | | | | | |
|---|---|---|---|---|---|---|---|---|
| Elastic Properties | C11 | | C12 | C13 | | C33 | C44 | |
| (GPa) | 143.0 | | 94.0 | 49.3 | | 191.0 | 18.0 | |
| Plastic Properties | $\dot{\gamma}_0$ | $m$ | $h_0$ | $\kappa_0^{\chi}$ | $\kappa_{s,0}$ | $\kappa_{s,S0}$ | $\dot{\gamma}_{S0}$ | $m'$ |
| | | | | (MPa) | | | | |
| Basal | 0.01 | 0.05 | 10 | 190 | 1 | 100 | $5\times10^{10}$ | 0.005 |
| Prismatic | 0.01 | 0.05 | 10 | 160 | 1 | 60 | $5\times10^{10}$ | 0.005 |
| Pyramidal | 0.01 | 0.05 | 10 | 400 | 1 | 420 | $5\times10^{10}$ | 0.005 |
| **$\beta$ phase Properties, $v_2$=0.06** | | | | | | | | |
| Elastic Properties | C11 | | C12 | C44 | | | | |
| (GPa) | 120.0 | | 108.0 | | 30.0 | | | |
| Plastic Properties | $\dot{\gamma}_0$ | $m$ | $h_0$ | $\kappa_0^{\chi}$ | $\kappa_{s,0}$ | $\kappa_{s,S0}$ | $\dot{\gamma}_{S0}$ | $m'$ |
| | | | (MPa) | | | | | |
| $\{110\}\langle111\rangle$ | 0.1 | 0.05 | 10 | 150 | 1 | 50 | 5x10$^{10}$ | 0.005 |
| $\{110\}\langle112\rangle$ | 0.1 | 0.05 | 10 | 170 | 1 | 75 | 5x10$^{10}$ | 0.005 |
| $\{110\}\langle123\rangle$ | 0.1 | 0.05 | 10 | 200 | 1 | 120 | 5x10$^{10}$ | 0.005 |
| **Void growth model parameters (for $\alpha$ and $\beta$ phases)** | | | | | | | | |
| $\mathcal{B}$ | $\mathcal{D}$ | $\mathcal{E}$ | $\mathcal{F}$ | $\mathcal{G}$ | $\mathcal{H}$ | $\mathcal{I}$ | $\mathcal{J}$ | |
| 1.20 | 0 | 5.30 | 1.20 | 1.25 | 7.00 | 1.80 | 2.50 | |
| **Porous plasticity parameters (for $\alpha$ and $\beta$ phases)** | | | | | | | | |
| $A_n$ | $\mathcal{s}_1$ | $\mathcal{s}_2$ | $\mathcal{s}_3$ (MPa) | $\xi_{gc}$ | $a_1$ | $a_2$ | $\xi_{crit}$ | |
| 0.02 | 0.018 | 0.16 | 70 | 0.16 | 10 | 1.1 | 0.65 | |

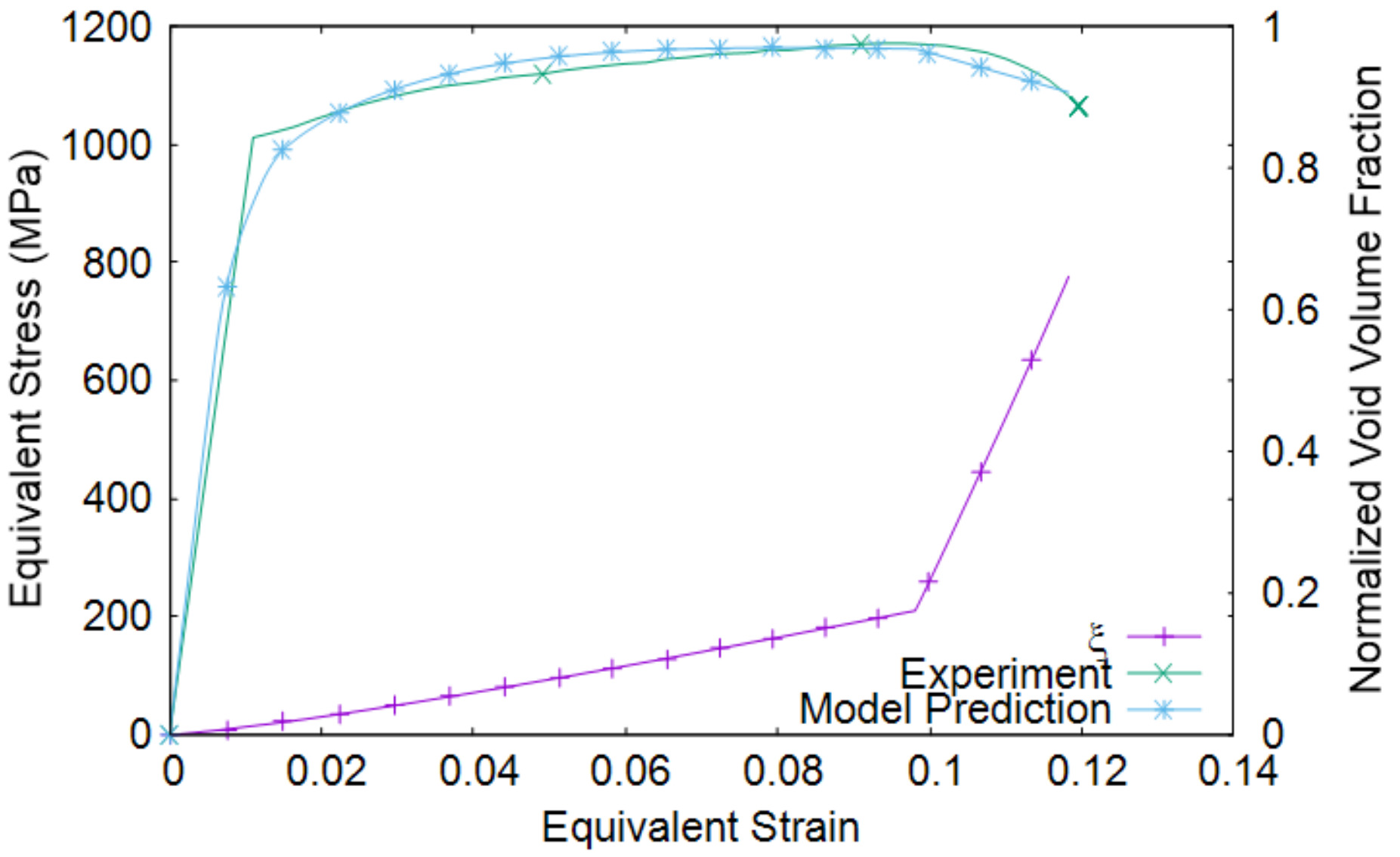

*Figure 5-1: Comparison of the constitutive model prediction with experimental results* [60]. *Predicted void growth is also shown.*

## 5.2 Forming limit curve simulations

The constitutive model presented in this work was calibrated using FLC results reported in the literature [60]. To replicate the FLC test, a single 3D cubic element is modelled utilising the material behaviour discussed in the previous sections. Three mutually perpendicular faces are constrained to remain stationary in the directions normal to those faces. Two opposite faces are assigned displacements normal to the faces such that:

$$\epsilon_2 = n\epsilon_1 \qquad (5\text{-}1)$$

Here $\epsilon_1$ and $\epsilon_2$ are the major and minor principal strains, respectively and $n$, no greater then unity, is a proportionality factor. The remaining face of the cube is left unconstrained and free of traction, representing the through-thickness behaviour of the thin sheet. The factor $n$ in equation (5-1) is also the slope of a line drawn from the origin to a point on an experimental FLC curve. The value of $n$ is set to be between -0.3 and 1.0 with the extremes representing uniaxial (-0.3) and biaxial (1.0) tension. Results are obtained at incremental values of $n$ equal to 0.1 in between. The values of $\epsilon_1$ and $\epsilon_2$ at which the $\xi_{crit}$ is achieved are then recorded to characterise the strain state at failure, and the resulting value of major strain is plotted in Figure 5-2 against the value of minor strain.

It has been reported in the literature that as the Lode parameter value is increased from 0 to 1 (right hand side of FLC) at higher values of stress triaxialities, void coalescence is delayed and failure is delayed to an even greater extent [41], [52], [53], [55]. To account for this effect of non-proportional straining of material points, a phenomenological relation for $\xi_{gc}$ is proposed. In this relationship, for $L_M$ greater than 0, the value of $\xi_{gc}$ is replaced by $\xi'_{gc}$, where $\xi'_{gc}$ is a function of the macroscopic value of the Lode parameter, $L_M$ and is given by:

$$\xi'_{gc} = \mathcal{g}_1 \xi_{gc} \exp\big(\mathcal{g}_2(L_M - 0.45)\big), \qquad L_M > 0 \tag{5-2}$$

Here, $\mathcal{g}_1$ and $\mathcal{g}_2$ are the material parameters which are to be calibrated with the experimental FLC results. The parameter $\xi_{crit}$ is needed in this application as the value of $\xi$ at which a material fails, and the element is deleted. The values of material parameters used in the void growth and coalescence part of the constitutive model were then found iteratively by an inverse modelling approach until the model predictions are in agreement with experimental FLC results from the literature, and are given in Table 5-2. Comparison of the two FLCs is given in Figure 5-2 which shows good agreement between the experimental FLC and the model prediction.

*Table 5-2: Void growth and coalescence parameters for the constitutive model calibrated from FLC results from Ti-6Al-4V*

| **Void growth model parameters (for $\alpha$ and $\beta$ phases)** | | | | | | | |
|---|---|---|---|---|---|---|---|
| $\mathcal{B}$ | $\mathcal{D}$ | $\mathcal{E}$ | $\mathcal{F}$ | $\mathcal{G}$ | $\mathcal{H}$ | $\mathcal{I}$ | $\mathcal{J}$ |
| 1.20 | 5.00 | 5.00 | 1.20 | 1.25 | 1.00 | 1.80 | 2.50 |
| **Porous plasticity parameters (for $\alpha$ and $\beta$ phases)** | | | | | | | |
| $A_n$ | $\mathcal{s}_1$ | $\mathcal{s}_2$ | $\mathcal{s}_3$ (MPa) | $\xi_{gc}$ | $\mathcal{g}_1$ | $\mathcal{g}_2$ | $\xi_{crit}$ |
| 4.5x10$^{-5}$ | 0.05 | 0.16 | 70.00 | 0.6 | 0.70 | 2.50 | 0.601 |

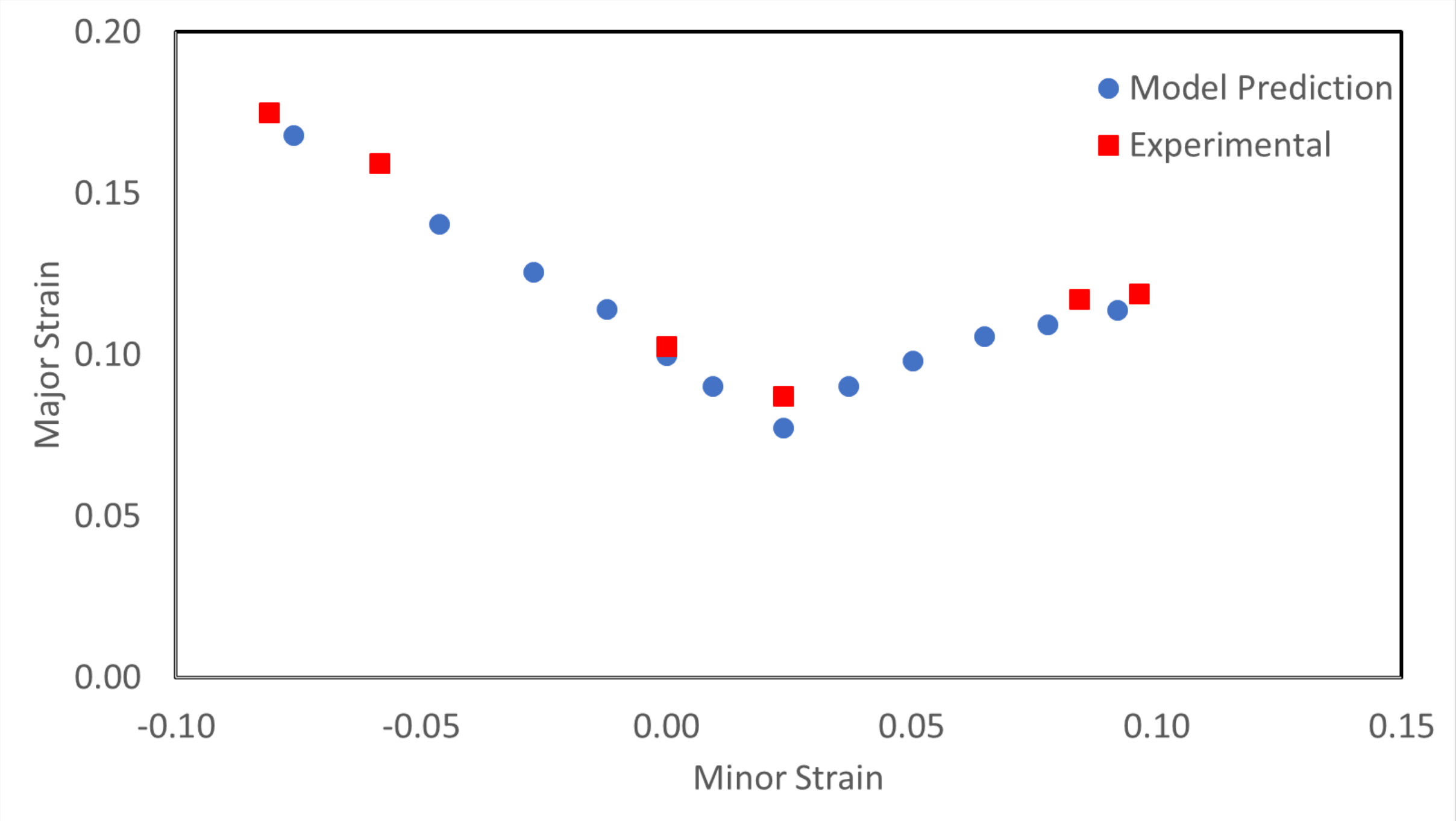

*Figure 5-2: The comparison of the constitutive model prediction and the experimental FLC curve of Ti-6Al-4V* [60]

# 6 Conclusion

A new constitutive model based on a crystal plasticity formulation is presented for dual phase titanium alloys which captures all stages of deformation, i.e. elasticity, plasticity, and ductile failure, under different loading conditions. The results of a CPFEM RVE study of void growth on the interface between the $\alpha$ and $\beta$ phases of a titanium alloy (Ti-10V-2Fe-3Al) reported recently are used to develop a void nucleation, growth and coalescence model. This model takes into account the effects of deformation state (equivalent strain), stress triaxiality, the Lode parameter and phase boundary inclination (PBI) on void growth. The proposed model is then incorporated in the formulation to extend its capability to cater for void nucleation, growth and coalescence. The resulting dual phase crystal plasticity model is then implemented as a user-subroutine in a commercially available finite element solver. An extensive parametric study has been carried out to explore the effects of material parameters used in the formulation. The performance of the model for various stress triaxialities and PBIs is also presented. The verification of the model is carried out by comparing the results of the RVE study of void growth at the interface of the $\alpha$ and $\beta$ phases of a titanium alloy and the model predictions, which showed good agreement. Application of the model is then demonstrated by simulating the experimental forming limit curve (FLC) for an $\alpha - \beta$ titanium alloy. The constitutive model calibrated to an $\alpha - \beta$ titanium alloy can be utilised for the design and optimisation of metal forming processes for dual phase titanium and similar alloys.

## Acknowledgement

This work was supported through a University of Aberdeen Elphinstone Scholarship which covered the tuition fee for PhD study.

## Appendices

## A. Crystal plasticity model with void growth for metals with single crystals that are almost elastically isotropic, have infinitesimal elastic strain and remain isotropic during void growth

The contribution of slip to the rate of deformation is given by:

$$\boldsymbol{D}_s^p = \sum_{\chi=1}^{N} \dot{\gamma}^{\chi} \mathrm{sym}(\boldsymbol{Z}^{\chi}) \tag{A-1}$$

where $\boldsymbol{Z}^{\chi} = \boldsymbol{s}^{\chi} \otimes \boldsymbol{m}^{\chi}$, is the Schmid tensor. And the void growth contribution is:

$$\boldsymbol{D}_v^p = \frac{1}{3} A_n \dot{\xi} \boldsymbol{I} \tag{A-2}$$

The total rate of deformation is given by:

$$\boldsymbol{d} = \frac{1}{2}(\boldsymbol{l} + \boldsymbol{l}^T) \tag{A-3}$$

Here $\boldsymbol{l} = \dot{\boldsymbol{F}}\boldsymbol{F}^{-1}$, is the total velocity gradient. The total rate of deformation is the sum of slip and void growth contributions plus an elastic part, which in terms of the Green-McInnis-Naghdi type of small elastic strain rate, $\overset{\nabla}{\boldsymbol{\epsilon}}{}^e$, is given by:

$$\boldsymbol{d} = \boldsymbol{D}_s^p + \boldsymbol{D}_v^p + \overset{\nabla}{\boldsymbol{\epsilon}}{}^e = \sum_{\chi=1}^{N} \dot{\gamma}^{\chi} \mathrm{sym}(\boldsymbol{Z}^{\chi}) + \frac{1}{3} A_n \dot{\xi} \boldsymbol{I} + \dot{\boldsymbol{\epsilon}}^e + \boldsymbol{\epsilon}^e \boldsymbol{w} - \boldsymbol{w} \boldsymbol{\epsilon}^e \tag{A-4}$$

where $\boldsymbol{w}$ is the spin tensor given by:

$$\boldsymbol{w} = \frac{1}{2}(\boldsymbol{l} - \boldsymbol{l}^T) \tag{A-5}$$

The dilatational rate is given by:

$$d_{kk} = \mathrm{tr}(\boldsymbol{d}) = \mathrm{tr}\left(\overset{\nabla}{\boldsymbol{\epsilon}}^{\boldsymbol{e}}\right) + A_n\dot{\xi} \tag{A-6}$$

so that the plastic part is

$$d_{kk}^p = \mathrm{tr}(\boldsymbol{D}^p) = A_n\dot{\xi} \tag{A-7}$$

## A.1 Elasticity

Elasticity is modelled as isotropic. Since elastic strain is infinitesimal and the elasticity is isotropic, we can write the elasticity relationship as:

$$\overset{\nabla}{\boldsymbol{\sigma}} = \mathbb{M}\colon \overset{\nabla}{\boldsymbol{\epsilon}}^e \tag{A-8}$$

where $\boldsymbol{\sigma}$ is Cauchy stress, $\overset{\nabla}{\boldsymbol{\sigma}}$ is the Green-McInnis-Naghdi rate of change of Cauchy stress, given by:

$$\overset{\nabla}{\boldsymbol{\sigma}} = \dot{\boldsymbol{\sigma}} + \boldsymbol{w\sigma} - \boldsymbol{\sigma w} \tag{A-9}$$

and the isotropic elasticity tensor is $\mathbb{M}$, which can be a function of the normalised void volume fraction, $\xi$.

Therefore, the constitutive law can be written as:

$$\mathbb{M}\colon \boldsymbol{d} = \sum_{\chi=1}^{N} \dot{\gamma}^{\chi}\big(\mathbb{M}\colon \mathrm{sym}(\boldsymbol{Z}^{\chi})\big) + \frac{1}{3}A_n\dot{\xi}(\mathbb{M}\colon \boldsymbol{I}) + \dot{\boldsymbol{\sigma}} + \boldsymbol{w\sigma} - \boldsymbol{\sigma w} \tag{A-10}$$

or

$$\dot{\boldsymbol{\sigma}} = \mathbb{M}\colon \left[\boldsymbol{d} - \sum_{\chi=1}^{N} \dot{\gamma}^{\chi}\mathrm{sym}(\boldsymbol{Z}^{\chi}) - \frac{1}{3}A_n\dot{\xi}\boldsymbol{I}\right] - \boldsymbol{w\sigma} + \boldsymbol{\sigma w} \tag{A-11}$$

## A.2 Update of the slip system rotation

The easiest way of seeing what to do is to investigate the case where the shear and bulk moduli are infinite so that the elastic strain is zero. In that case

$$\boldsymbol{F}^p = \boldsymbol{F} \tag{A-12}$$

and

$$\dot{\boldsymbol{F}} = \boldsymbol{lF} = (\boldsymbol{d} + \boldsymbol{w})\boldsymbol{F} = \left(\boldsymbol{D}_s^p + \boldsymbol{D}_v^p + \boldsymbol{w}\right)\boldsymbol{F} \tag{A-13}$$

allowing integration of $\boldsymbol{F}$ with respect to time.

This can also be addressed for the case when there is elastic strain with finite shear and bulk moduli. The total deformation gradient can be multiplicatively decomposed into elastic and plastic parts:

$$\boldsymbol{F} = \boldsymbol{F}^e\boldsymbol{F}^P \tag{A-14}$$

The elastic part can be written as:

$$\boldsymbol{F}^e = \boldsymbol{I} + \boldsymbol{\epsilon}^e - \boldsymbol{\omega}^e \tag{A-15}$$

where $\boldsymbol{\epsilon}^e$ is symmetric and $\boldsymbol{\omega}^e$ is skew, and contain terms that are all small in magnitude compared to unity.

Therefore to 1st order:

$$\boldsymbol{F} = \boldsymbol{F}^p \tag{A-16}$$

a result already given in (A-12). The result from integrating (A-13) will differ from an exact result only by 2nd order terms.

Thus (A-13) can be written as:

$$\dot{\boldsymbol{F}}^p = \left(\boldsymbol{D}_s^p + \boldsymbol{D}_v^p + \boldsymbol{w}\right)\boldsymbol{F}^p \tag{A-17}$$

which we can use to integrate $\dot{\boldsymbol{F}}^p$ with respect to time.

The final step, once $\boldsymbol{F}^p$ is calculated, is to use polar decomposition to obtain the plastic rotation.

# B. Numerical Implementation

## B.1 Anisotropic elasticity

The Kirchhoff stress given in (2-29) was split into deviatoric, $\text{dev}\boldsymbol{\tau}$ and volumetric parts, $p_\tau$, which in return relate to the deviatoric and volumetric parts of elastic strains, $\text{dev}\boldsymbol{\epsilon}^e$ and $\epsilon_{kk}^e$, respectively, using the following relations:

$$\begin{aligned} \text{dev}\boldsymbol{\tau} &= \tilde{\mathbb{C}}_d^e : \text{dev}\boldsymbol{\epsilon}^e + \widetilde{\boldsymbol{H}}^e \epsilon_{kk}^e \\ p_\tau &= \widetilde{\boldsymbol{H}}^{e^T} : \text{dev}\boldsymbol{\epsilon}^e + \widetilde{M}^e \epsilon_{kk}^e \end{aligned} \tag{B-1}$$

Here, $\tilde{\mathbb{C}}_d^e$, $\widetilde{\boldsymbol{H}}^e$ and $\widetilde{M}^e$ are the deviatoric fourth-order elastic tensor, deviatoric-isochoric elastic-coupling second-order tensor and elastic volumetric coefficient respectively, and are evaluated using:

$$\tilde{\mathbb{C}}_d^e = \widetilde{\mathbb{P}}_d : \tilde{\mathbb{C}}^e : \widetilde{\mathbb{P}}_d, \qquad \widetilde{\boldsymbol{H}}^e = \frac{1}{3}\widetilde{\mathbb{P}}_d : \tilde{\mathbb{C}}^e : \mathbf{1}, \qquad \widetilde{M}^e = \frac{1}{9}\mathbf{1} : \tilde{\mathbb{C}}^e : \mathbf{1} \tag{B-2}$$

Here, $\widetilde{\mathbb{P}}_d = \tilde{\mathbb{I}} - \frac{1}{3}\mathbf{1} \otimes \mathbf{1}$ and $\mathbf{1}$ is second-order identity tensor. Components of $\widetilde{\boldsymbol{H}}^e$ for crystals with cubic symmetry are zero but have finite values for other crystal symmetries.

## B.2 Constitutive integration scheme

Relations given in (2-28), (2-29) and (2-30) are a set of coupled first order ordinary differential equations in the variables $(\boldsymbol{\tau}, \xi, \boldsymbol{R}^e, \kappa_s^\chi)$ which are to be solved. First, the evolution equations of the deformation are discretised in time and then numerically integrated to get the results for each time step. The current time is represented as $t_n$ and the integration is carried out to get the results at time $t_{n+1}$, and are related as $t_{n+1} = t_n + \Delta t$. The values of quantities at time $t_n$ and $t_{n+1}$ are represented with subscripts $n$ and $n+1$, respectively. The following are needed for the integration to proceed:

i. Updated value of the deformation at $t_{n+1}$ in the form of $\boldsymbol{l}_{n+1}$ or $\boldsymbol{d}_{n+1}$ and $\boldsymbol{w}_{n+1}$
ii. The values of $\boldsymbol{\tau}_n, \boldsymbol{R}_n^e, \kappa_{s,n}^\chi$ and $\xi_n$
iii. Time-independent values of slip system unit vectors $\left(\boldsymbol{s}_0^\chi, \boldsymbol{m}_0^\chi\right)$ in the sample coordinate system, the initial orientation of the crystal in terms of its Euler angles from which the rotation matrix, $\boldsymbol{C}_0$ can be calculated, the elasticity tensor in the sample coordinate system $\mathbb{C}_0^e$ and the plasticity parameters required to solve the flow rule, hardening evolution, and void growth and coalescence.

Updated of values of $(\boldsymbol{\tau}_{n+1}, \xi_{n+1}, \boldsymbol{R}_{n+1}^e, \kappa_{s,n+1}^\chi)$ can then be calculated using the integration of the relevant set of equations.

The numerical integration scheme used in this work is as follows. The kinematic equation given in (2-28) can be written as:

$$\overset{\nabla}{\boldsymbol{\epsilon}}^e = \boldsymbol{R}^e \left[\frac{\partial}{\partial t}\left(\boldsymbol{R}^{e^T}\boldsymbol{\epsilon}^e\boldsymbol{R}^e\right)\right]\boldsymbol{R}^{e^T} = \boldsymbol{d} - \widetilde{\boldsymbol{D}}^p \tag{B-3}$$

This can be integrated using the backward Euler scheme as:

$$\boldsymbol{R}_{n+1}^{eT}\boldsymbol{\epsilon}_{n+1}^e\boldsymbol{R}_{n+1}^e = \boldsymbol{R}_n^{e^T}\boldsymbol{\epsilon}_n^e\boldsymbol{R}_n^e + \Delta t\boldsymbol{R}_{n+1}^{eT}(\boldsymbol{d}_{n+1} - \widetilde{\boldsymbol{D}}_{n+1}^p)\boldsymbol{R}_{n+1}^e \tag{B-4}$$

or

$$\boldsymbol{\epsilon}_{n+1}^e = \underbrace{\boldsymbol{R}_{n+1}^e\boldsymbol{R}_n^{e^T}}_{\Delta\boldsymbol{R}^e}\boldsymbol{\epsilon}_n^e\underbrace{\boldsymbol{R}_n^e\boldsymbol{R}_{n+1}^{eT}}_{\Delta\boldsymbol{R}^{eT}} + \Delta t(\boldsymbol{d}_{n+1} - \widetilde{\boldsymbol{D}}_{n+1}^p) \tag{B-5}$$

where $\Delta\boldsymbol{R}^e$ is the incremental elastic rotation tensor. The plastic strain rate, given in (2-30), can then be updated using:

$$\begin{aligned}\widetilde{\boldsymbol{D}}_{n+1}^p &= \boldsymbol{R}_{n+1}^e\overline{\boldsymbol{D}}_{n+1}^p\boldsymbol{R}_{n+1}^{eT} = \boldsymbol{R}_{n+1}^e\mathrm{sym}\left(\sum_{\chi=1}^{\mathrm{N}}\dot{\gamma}_{n+1}^\chi\ \overline{\boldsymbol{s}}^\chi\otimes\overline{\boldsymbol{m}}^\chi\right)\boldsymbol{R}_{n+1}^{eT} + \frac{1}{3}A_n\dot{\xi}_{n+1}\boldsymbol{1} \\ &= \sum_{\chi=1}^{\mathrm{N}}\dot{\gamma}_{n+1}^\chi\ \mathrm{sym}\left(\underbrace{\boldsymbol{R}_{n+1}^e\boldsymbol{C}_0}_{\boldsymbol{C}_{n+1}}\ \boldsymbol{s}_0^\chi\otimes\underbrace{\boldsymbol{R}_{n+1}^e\boldsymbol{C}_0}_{\boldsymbol{C}_{n+1}}\boldsymbol{m}_0^\chi\right) + \frac{1}{3}A_n\dot{\xi}_{n+1}\boldsymbol{1} \\ &= \sum_{\chi=1}^{\mathrm{N}}\dot{\gamma}_{n+1}^\chi\ \mathrm{sym}\left(\widetilde{\boldsymbol{s}}_{n+1}^\chi\otimes\widetilde{\boldsymbol{m}}_{n+1}^\chi\right) + \frac{1}{3}A_n\dot{\xi}_{n+1}\boldsymbol{1}\end{aligned} \tag{B-6}$$

The updated value of $\boldsymbol{R}_{n+1}^e$ is calculated using the exponential map [61], utilizing (2-28) and (2-30), and can be written as:

$$\boldsymbol{R}_{n+1}^e = \exp\left(\Delta t\widetilde{\boldsymbol{\Omega}}_{n+1}^e\right)\boldsymbol{R}_n^e, \qquad \widetilde{\boldsymbol{\Omega}}_{n+1}^e = \boldsymbol{w}_{n+1} - \sum_{\chi=1}^{\mathrm{N}}\dot{\gamma}_{n+1}^\chi\ \mathrm{skew}\left(\widetilde{\boldsymbol{s}}_{n+1}^\chi\otimes\widetilde{\boldsymbol{m}}_{n+1}^\chi\right) \tag{B-7}$$

In (B-6) and (B-7), the value of $\dot{\gamma}_{n+1}^\chi$ is calculated using the flow rule given in (2-35). Updated slip system vectors $\left(\widetilde{\boldsymbol{s}}_{n+1}^\chi, \widetilde{\boldsymbol{m}}_{n+1}^\chi\right)$ are evaluated using $\widetilde{\boldsymbol{s}}_{n+1}^\chi = \boldsymbol{C}_{n+1}\boldsymbol{s}_0^\chi$ and $\widetilde{\boldsymbol{m}}_{n+1}^\chi = \boldsymbol{C}_{n+1}\boldsymbol{m}_0^\chi$, where $\boldsymbol{C}_{n+1} = \boldsymbol{R}_{n+1}^e\boldsymbol{C}_0$ is the updated rotation tensor. The value of $\dot{\xi}_{n+1}$ is calculated using (2-31) - (2-34), depending on the value of $\xi_n$. The value of $\mathrm{pbi}_{n+1}$ in (2-32) is also updated using $\boldsymbol{R}_{n+1}^e$. The symmetric and skew parts of the Schmid tensor in the current configuration and at time $t_{n+1}$ are represented as:

$$\widetilde{\boldsymbol{P}}_{n+1}^\chi = \mathrm{sym}\left(\widetilde{\boldsymbol{s}}_{n+1}^\chi\otimes\widetilde{\boldsymbol{m}}_{n+1}^\chi\right), \qquad \widetilde{\boldsymbol{Q}}_{n+1}^\chi = \mathrm{skew}\left(\widetilde{\boldsymbol{s}}_{n+1}^\chi\otimes\widetilde{\boldsymbol{m}}_{n+1}^\chi\right) \tag{B-8}$$

The elasticity tensors can be rotated to the current configuration using:

$$\widetilde{\mathbb{C}}_{n+1}^e = (\boldsymbol{C}_{n+1}\otimes\boldsymbol{C}_{n+1}):\widetilde{\mathbb{C}}_0^e:(\boldsymbol{C}_{n+1}\otimes\boldsymbol{C}_{n+1})^T, \qquad \widetilde{H}_{n+1}^e = \boldsymbol{C}_{n+1}\widetilde{H}_0^e\boldsymbol{C}_{n+1}^T \tag{B-9}$$

The elastic strains at time $t_{n+1}$ can then be written using (B-6), as:

$$\boldsymbol{\epsilon}_{n+1}^e = \underbrace{\hat{\boldsymbol{\epsilon}}_{n+1}^e + \Delta t\boldsymbol{d}_{n+1}}_{\boldsymbol{\epsilon}_{n+1}^{e*}} - \Delta t\sum_{\chi=1}^{\mathrm{N}}\dot{\gamma}_{n+1}^\chi\ \widetilde{\boldsymbol{P}}_{n+1}^\chi - \Delta t\frac{1}{3}A_n\dot{\xi}_{n+1}\boldsymbol{1}, \qquad \hat{\boldsymbol{\epsilon}}_{n+1}^e = \Delta\boldsymbol{R}^e\boldsymbol{\epsilon}_n^e\Delta\boldsymbol{R}^{eT} \tag{B-10}$$

and can thereafter be expressed in deviatoric and volumetric parts as:

$$\mathrm{dev}\boldsymbol{\epsilon}_{n+1}^e = \mathrm{dev}\boldsymbol{\epsilon}_{n+1}^{e*} - \Delta t\sum_{\chi=1}^{\mathrm{N}}\dot{\gamma}_{n+1}^\chi\ \widetilde{\boldsymbol{P}}_{n+1}^\chi, \qquad \epsilon_{kk,n+1}^e = \epsilon_{kk,n+1}^{e*} - \Delta tA_n\dot{\xi}_{n+1} \tag{B-11}$$

where $\mathrm{dev}\boldsymbol{\epsilon}_{n+1}^{e*}$ and $\epsilon_{kk,n+1}^{e*}$ are elastic predictor deviatoric and volumetric strains respectively, given by:

$$\mathrm{dev}\boldsymbol{\epsilon}_{n+1}^{e*} = \Delta\boldsymbol{R}^e\mathrm{dev}\boldsymbol{\epsilon}_n^e\Delta\boldsymbol{R}^{eT} + \Delta t\mathrm{dev}\boldsymbol{d}_{n+1}, \qquad \epsilon_{kk,n+1}^{e*} = \epsilon_{kk,n}^e + \Delta td_{kk,n+1} \tag{B-12}$$

Here, $d_{kk,n+1} = \mathrm{tr}(\boldsymbol{d}_{n+1})$ is the volumetric part of the rate of deformation tensor. The relations given in (B-11) and (B-12) can then be used in (B-1) and rearranged to get the values at time $t_{n+1}$:

$$\tilde{\mathbb{C}}^{e-1}_{d,n+1}:\text{dev}\boldsymbol{\tau}_{n+1} = \text{dev}\boldsymbol{\epsilon}^{e}_{n+1} + \tilde{\mathbb{C}}^{e-1}_{d,n+1}:\widetilde{\boldsymbol{H}}^{e}_{n+1}\epsilon^{e}_{kk,n+1}$$
$$\widetilde{M}^{e-1}_{n+1}p_{\tau,n+1} = \widetilde{M}^{e-1}_{n+1}\left(\widetilde{\boldsymbol{H}}^{eT}_{n+1}:\text{dev}\boldsymbol{\epsilon}^{e}_{n+1}\right) + \epsilon^{e}_{kk,n+1} \tag{B-13}$$

to obtain:

$$\tilde{\mathbb{C}}^{e-1}_{d,n+1}:\text{dev}\boldsymbol{\tau}_{n+1} = \text{dev}\boldsymbol{\epsilon}^{e*}_{n+1} - \Delta t\sum_{\chi=1}^{\text{N}}\dot{\gamma}^{\chi}_{n+1}\widetilde{\boldsymbol{P}}^{\chi}_{n+1} + \tilde{\mathbb{C}}^{e-1}_{d,n+1}:\widetilde{\boldsymbol{H}}^{e}_{n+1}\epsilon^{e}_{kk,n+1}$$
$$\widetilde{M}^{e-1}_{n+1}p_{\tau,n+1} = \widetilde{M}^{e-1}_{n+1}\left(\widetilde{\boldsymbol{H}}^{eT}_{n+1}:\text{dev}\boldsymbol{\epsilon}^{e}_{n+1}\right) + \epsilon^{e*}_{kk,n+1} - A_n\xi_{n+1} \tag{B-14}$$

The backward Euler scheme is used to calculate $\kappa^{\chi}_{s,n+1}$ at time $t_{n+1}$ using (2-37) and is written as:

$$\kappa^{\chi}_{s,n+1} = \kappa^{\chi}_{s,n} + \Delta t h_0\left(\frac{\kappa_{s,S,n+1} - \kappa^{\chi}_{s,n+1}}{\kappa_{s,S,n+1} - \kappa_{s,0}}\right)\sum_{\chi=1}^{N}|\dot{\gamma}_{n+1}{}^{\chi}| \tag{B-15}$$

The relations in (B-7), (B-14) (a-b) and (B-15) make a set of coupled algebraic equations to be solved to obtain the values of $\left(\text{dev}\boldsymbol{\tau}_{n+1}, p_{\tau,n+1}, \boldsymbol{R}^{e}_{n+1}, \kappa^{\chi}_{s,n+1}\right)$. The residuals can then be written using these equations, as:

$$\begin{aligned}\mathcal{R}_1 &= \hat{\mathcal{R}}_1\left(\text{dev}\boldsymbol{\tau}_{n+1}, p_{\tau,n+1}, \boldsymbol{R}^{e}_{n+1}, \kappa^{\chi}_{s,n+1}\right)\\ &= \tilde{\mathbb{C}}^{e-1}_{d,n+1}:\text{dev}\boldsymbol{\tau}_{n+1} - \text{dev}\boldsymbol{\epsilon}^{e*}_{n+1} + \Delta t\sum_{\chi=1}^{\text{N}}\dot{\gamma}^{\chi}_{n+1}\widetilde{\boldsymbol{P}}^{\chi}_{n+1}\\ &\quad - \tilde{\mathbb{C}}^{e-1}_{d,n+1}:\widetilde{\boldsymbol{H}}^{e}_{n+1}\epsilon^{e}_{kk,n+1} = 0\end{aligned} \tag{B-16}$$

$$\begin{aligned}\mathcal{R}_2 &= \hat{\mathcal{R}}_2\left(\text{dev}\boldsymbol{\tau}_{n+1}, p_{\tau,n+1}, \boldsymbol{R}^{e}_{n+1}, \kappa^{\chi}_{s,n+1}\right)\\ &= p_{\tau,n+1}\widetilde{M}^{e-1}_{n+1} - \epsilon^{e*}_{kk,n+1} + A_n\xi_{n+1} - \widetilde{M}^{e-1}_{n+1}\left(\widetilde{\boldsymbol{H}}^{eT}_{n+1}:\text{dev}\boldsymbol{\epsilon}^{e}_{n+1}\right) = 0\end{aligned} \tag{B-17}$$

$$\begin{aligned}\mathcal{R}_3 &= \hat{\mathcal{R}}_3\left(\text{dev}\boldsymbol{\tau}_{n+1}, p_{\tau,n+1}, \boldsymbol{R}^{e}_{n+1}, \kappa^{\chi}_{s,n+1}\right)\\ &= \boldsymbol{R}^{e}_{n+1} - exp\left[\Delta t\left(\boldsymbol{w}_{n+1} - \sum_{\chi=1}^{N}\dot{\gamma}^{\chi}_{n+1}\widetilde{\boldsymbol{Q}}^{\chi}_{n+1}\right)\right]\boldsymbol{R}^{e}_{n} = 0\end{aligned} \tag{B-18}$$

$$\begin{aligned}\mathcal{R}_4 &= \hat{\mathcal{R}}_4\left(\text{dev}\boldsymbol{\tau}_{n+1}, p_{\tau,n+1}, \boldsymbol{R}^{e}_{n+1}, \kappa^{\chi}_{s,n+1}\right)\\ &= \kappa^{\chi}_{s,n+1} - \kappa^{\chi}_{s,n} - \Delta t h_0\left(\frac{\kappa_{s,S,n+1} - \kappa^{\chi}_{s,n+1}}{\kappa_{s,S,n+1} - \kappa_{s,0}}\right)\sum_{\chi=1}^{N}|\dot{\gamma}_{n+1}{}^{\chi}| = 0\end{aligned} \tag{B-19}$$

A two-level staggered iterative scheme used in the formulation of Marin [56], who based his on the work of others [62], [63], is used in this work. In this scheme, the N-R method is used to solve the set of equations given in (B-16) and (B-17) for the residuals, in order to get the values of $\text{dev}\boldsymbol{\tau}_{n+1}$ and $p_{\tau,n+1}$ starting from the best estimates of $\left(\boldsymbol{R}^{e}_{n+1}, \kappa^{\alpha}_{s,n+1}\right)$. A linearisation of the residual given in (B-16) with respect to $\text{dev}\boldsymbol{\tau}$, gives rise to a set of algebraic equations which can be solved iteratively to get $\Delta(\text{dev}\boldsymbol{\tau}_{n+1})$ at a given time step using:

$$\begin{aligned}&\left(\tilde{\mathbb{C}}^{e-1}_{d,n+1} + \Delta t\sum_{\chi=1}^{\text{N}}\frac{\partial\dot{\gamma}^{\chi}_{n+1}}{\partial\tau^{\chi}_{n+1}}\widetilde{\boldsymbol{P}}^{\chi}_{n+1}\otimes\widetilde{\boldsymbol{P}}^{\chi}_{n+1}\right):\Delta(\text{dev}\boldsymbol{\tau}_{n+1})\\ &\qquad = -\text{dev}\boldsymbol{\epsilon}^{e}_{n+1} + \text{dev}\boldsymbol{\epsilon}^{e*}_{n+1} - \Delta t\sum_{\chi=1}^{\text{N}}\dot{\gamma}^{\chi}_{n+1}\widetilde{\boldsymbol{P}}^{\chi}_{n+1}\end{aligned} \tag{B-20}$$

Here, the relation given in (B-13) (a) is used to obtain $\text{dev}\boldsymbol{\epsilon}^{e}_{n+1} = \tilde{\mathbb{C}}^{e-1}_{d,n+1}:\left(\text{dev}\boldsymbol{\tau}_{n+1} - \widetilde{\boldsymbol{H}}^{e}_{n+1}\epsilon^{e}_{kk,n+1}\right)$. Based on the value of $\text{dev}\boldsymbol{\epsilon}^{e}_{n+1}$ just obtained, the N-R method is used to solve for the residual given in (B-17) iteratively for each time step to get $\Delta p_{\tau,n+1}$, using:

$$\left(\widetilde{M}^{e-1}_{n+1} + \frac{\partial\xi_{n+1}}{\partial p_{\tau,n+1}}\right)\Delta p_{\tau,n+1} = -\epsilon^{e}_{kk,n+1} + \epsilon^{e*}_{kk,n+1} - A_n\xi_{n+1} \tag{B-21}$$

where the relation given in (B-13) (b) is used to obtain $\epsilon^e_{kk,n+1} = \widetilde{M}^{e-1}_{n+1}\left(p_{\tau,n+1} - \widetilde{\boldsymbol{H}}^{eT}_{n+1}\colon \mathrm{dev}\boldsymbol{\epsilon}^e_{n+1}\right)$. After getting the values of $\mathrm{dev}\boldsymbol{\tau}_{n+1}$ and $p_{\tau,n+1}$, a second-level of the N-R scheme is used to calculate:

i. the value of $\kappa^{\chi}_{s,n+1}$ using (B-19) while keeping the values of $(\mathrm{dev}\boldsymbol{\tau}_{n+1}, \boldsymbol{R}^e_{n+1})$ constant

ii. and $\boldsymbol{R}^e_{n+1}$ using (B-18).

Details about the second-level of the iterative scheme can be found in the literature [62].

The value of Cauchy stress can then be calculated using:

$$\boldsymbol{\sigma}_{n+1} = \det(1 + \boldsymbol{\epsilon}^e_{n+1})\,\boldsymbol{\tau}_{n+1}, \qquad \boldsymbol{\tau}_{n+1} = \mathrm{dev}\boldsymbol{\tau}_{n+1} + p_{\tau,n+1}\mathbf{1} \tag{B-22}$$

by combining the deviatoric and volumetric parts of Kirchhoff stress. Also, the value of the total elastic strain is calculated for the time $t_{n+1}$ using:

$$\boldsymbol{\epsilon}^e_{n+1} = \mathrm{dev}\boldsymbol{\epsilon}^e_{n+1} + \frac{1}{3}\epsilon^e_{kk,n+1}\mathbf{1} \tag{B-23}$$

## B.3 Consistent elastoplastic tangent moduli

The finite element method we used takes advantage of the material tangent moduli for convergence while solving for equilibrium using an implicit scheme. An effort is made to find approximate tangent moduli using the constitutive equations. Since the derivation does not consider linearisation of the rotation tensor, $\boldsymbol{R}^e_{n+1}$, elastic Jacobian, $J^e_{n+1} = \det(1 + \boldsymbol{\epsilon}^e_{n+1})$ and the hardness, $\kappa^{\chi}_{s,n+1}$, it is termed an approximate modulus instead of an exact one.

The elastoplastic tangent moduli can then be defined as:

$$\boldsymbol{c}^{ep}_{n+1} = \frac{1}{\Delta t}\frac{\mathrm{d}\boldsymbol{\sigma}_{n+1}}{\mathrm{d}\boldsymbol{d}_{n+1}} = \frac{1}{J^e_{n+1}\Delta t}\frac{\mathrm{d}\boldsymbol{\tau}_{n+1}}{\mathrm{d}\boldsymbol{d}_{n+1}} \rightarrow \mathrm{d}\boldsymbol{\tau}_{n+1} = \underbrace{J^e_{n+1}\boldsymbol{c}^{ep}_{n+1}}_{\boldsymbol{c}^{ep}_{\tau,n+1}}\colon \mathrm{d}\boldsymbol{d}_{n+1}\Delta t \tag{B-24}$$

Here, $\boldsymbol{c}^{ep}_{n+1}$ are elastoplastic tangent moduli in terms of Cauchy stress, $\boldsymbol{\sigma}_{n+1}$ which can be transformed into Kirchhoff stress, $\boldsymbol{\tau}_{n+1}$ using $\boldsymbol{\tau}_{n+1} = J^e_{n+1}\boldsymbol{\sigma}_{n+1}$ and $\boldsymbol{c}^{ep}_{\tau,n+1}$ are the elastoplastic tangent moduli in terms of Kirchhoff stress.

The decomposition of $\boldsymbol{\tau}_{n+1}$ into deviatoric and volumetric parts is given by:

$$\boldsymbol{\tau}_{n+1} = \mathrm{dev}\boldsymbol{\tau}_{n+1} + p_{\tau,n+1}\mathbf{1} \tag{B-25}$$

Then using (B-1), constitutive equations can be written in time $t_{n+1}$:

$$\begin{aligned} \mathrm{dev}\boldsymbol{\tau}_{n+1} &= \widetilde{\mathbb{C}}^e_{d,n+1}\colon \mathrm{dev}\boldsymbol{\epsilon}^e_{n+1} + \widetilde{\boldsymbol{H}}^e_{n+1}\epsilon^e_{kk,n+1} \\ p_{\tau,n+1} &= \widetilde{\boldsymbol{H}}^{eT}_{n+1}\colon \mathrm{dev}\boldsymbol{\epsilon}^e_{n+1} + \widetilde{M}^{e-1}_{n+1}\epsilon^e_{kk,n+1} \end{aligned} \tag{B-26}$$

Here, the elastic strain is used in the form of deviatoric and volumetric parts, which are given by:

$$\begin{aligned} \mathrm{dev}\boldsymbol{\epsilon}^e_{n+1} &= \mathrm{dev}\hat{\boldsymbol{\epsilon}}^e_{n+1} + \Delta t\,\mathrm{dev}\boldsymbol{d}_{n+1} - \Delta t\sum_{\chi=1}^{\mathrm{N}}\dot{\gamma}^{\chi}_{n+1}\,\widetilde{\boldsymbol{P}}^{\chi}_{n+1} \\ \epsilon^e_{kk,n+1} &= \hat{\epsilon}^e_{kk,n+1} + \Delta t\,d_{kk,n+1} - A_n\xi_{n+1} \end{aligned} \tag{B-27}$$

where $\hat{\epsilon}^e_{kk,n+1} = \epsilon^e_{kk,n}$, since it is a scalar quantity and is not affected by rotation. A linearisation of $\boldsymbol{\tau}_{n+1}$ in (B-25) leads to:

$$\mathrm{d}\boldsymbol{\tau}_{n+1} = \mathrm{d}\,\mathrm{dev}\boldsymbol{\tau}_{n+1} + \mathrm{d}p_{\tau,n+1}\mathbf{1} \tag{B-28}$$

Since our approximation is that $\boldsymbol{R}^e_{n+1}$ is constant, values of $\widetilde{\mathbb{C}}^{e-1}_{d,n+1}$, $\widetilde{\boldsymbol{H}}^e_{n+1}$, $\widetilde{M}^{e-1}_{n+1}$ and $\widetilde{\boldsymbol{P}}^{\chi}_{n+1}$ will be treated as constants as well. Now $\mathrm{dev}\boldsymbol{\tau}_{n+1}$ and $p_{\tau,n+1}$ in (B-26), can be linearised using:

$$\widetilde{\mathbb{C}}^{e-1}_{d,n+1}\colon \mathrm{d}\,\mathrm{dev}\boldsymbol{\tau}_{n+1} = \mathrm{d}\,\mathrm{dev}\boldsymbol{\epsilon}^e_{n+1} + \widetilde{\mathbb{C}}^{e-1}_{d,n+1}\colon \widetilde{\boldsymbol{H}}^e_{n+1}\mathrm{d}\epsilon^e_{kk,n+1} \tag{B-29}$$

where $\mathrm{d}\,\mathrm{dev}\boldsymbol{\epsilon}^e_{n+1}$ can be written, using (B-27) (a) as:

$$\mathrm{d}\ \mathrm{dev}\boldsymbol{\epsilon}_{n+1}^{e} = \Delta t\widetilde{\mathbb{P}}_d : \mathrm{d}\boldsymbol{d}_{n+1} - \underbrace{\sum_{\chi=1}^{N} \Delta t \frac{\partial \dot{\gamma}_{n+1}^{\chi}}{\partial \tau_{n+1}^{\chi}} \left(\widetilde{\boldsymbol{P}}_{n+1}^{\chi} \otimes \widetilde{\boldsymbol{P}}_{n+1}^{\chi}\right)}_{\widetilde{\mathbb{S}}_{n+1}} : \mathrm{d}\ \mathrm{dev}\boldsymbol{\tau}_{n+1} \tag{B-30}$$

Similarly, the volumetric part can be written as:

$$\widetilde{M}_{n+1}^{e-1}\mathrm{d}p_{\tau,n+1} = \widetilde{M}_{n+1}^{e-1}\left(\widetilde{\boldsymbol{H}}_{n+1}^{eT} : \mathrm{d}\ \mathrm{dev}\boldsymbol{\epsilon}_{n+1}^{e}\right) + \mathrm{d}\epsilon_{kk,n+1}^{e} \tag{B-31}$$

Using (B-27) (b):

$$\mathrm{d}\epsilon_{kk,n+1}^{e} = \Delta t\boldsymbol{1} : \mathrm{d}\boldsymbol{d}_{n+1} - A_n \frac{\partial \xi_{n+1}}{\partial p_{\tau,n+1}} \mathrm{d}p_{\tau,n+1} \tag{B-32}$$

The values of $\mathrm{d}\ \mathrm{dev}\boldsymbol{\epsilon}_{n+1}^{e}$ and $\mathrm{d}\epsilon_{kk,n+1}^{e}$ can then be substituted in (B-29) from (B-30) and (B-32), to get:

$$\begin{aligned} \widetilde{\mathbb{C}}_{d,n+1}^{e-1} : \mathrm{d}\ \mathrm{dev}\boldsymbol{\tau}_{n+1} &= \Delta t\widetilde{\mathbb{P}}_d : \mathrm{d}\boldsymbol{d}_{n+1} - \widetilde{\mathbb{S}}_{n+1} : \mathrm{d}\ \mathrm{dev}\boldsymbol{\tau}_{n+1} \\ &+ \widetilde{\mathbb{C}}_{d,n+1}^{e-1} : \widetilde{\boldsymbol{H}}_{n+1}^{e} \left(\Delta t\boldsymbol{1} : \mathrm{d}\boldsymbol{d}_{n+1} - A_n \frac{\partial \xi_{n+1}}{\partial p_{\tau,n+1}} \mathrm{d}p_{\tau,n+1}\right) \end{aligned} \tag{B-33}$$

and in (B-31), getting:

$$\begin{aligned} \widetilde{M}_{n+1}^{e-1}\mathrm{d}p_{\tau,n+1} = \widetilde{M}_{n+1}^{e-1}\left\{\widetilde{\boldsymbol{H}}_{n+1}^{eT} : \left(\Delta t\widetilde{\mathbb{P}}_d : \mathrm{d}\boldsymbol{d}_{n+1} - \widetilde{\mathbb{S}}_{n+1} : \mathrm{d}\ \mathrm{dev}\boldsymbol{\tau}_{n+1}\right)\right\} + \Delta t\boldsymbol{1} : \mathrm{d}\boldsymbol{d}_{n+1} \\ - A_n \frac{\partial \xi_{n+1}}{\partial p_{\tau,n+1}} \mathrm{d}p_{\tau,n+1} \end{aligned} \tag{B-34}$$

The relations given in (B-33) and (B-34) can then be rearranged and written as a system of equations which can be solved for $\mathrm{d}\ \mathrm{dev}\boldsymbol{\tau}_{n+1}$ and $\mathrm{d}p_{\tau,n+1}$ in terms of $\mathrm{d}\boldsymbol{d}_{n+1}\Delta t$. Their *coefficients* (not strictly scalar, can be tensors) $\mathbb{G}_1$, $\boldsymbol{G}_2$, $\mathbb{G}_3$, $\boldsymbol{G}_4$, $G_5$ and $\boldsymbol{G}_6$ are described here as:

$$\begin{aligned} \underbrace{\left(\widetilde{\mathbb{C}}_{d,n+1}^{e-1} + \widetilde{\mathbb{S}}_{n+1}\right)}_{\mathbb{G}_1} : \mathrm{d}\ \mathrm{dev}\boldsymbol{\tau}_{n+1} + \underbrace{\widetilde{\mathbb{C}}_{d,n+1}^{e-1} : \widetilde{\boldsymbol{H}}_{n+1}^{e} \left(A_n \frac{\partial \xi_{n+1}}{\partial p_{\tau,n+1}}\right)}_{\boldsymbol{G}_2} \mathrm{d}p_{\tau,n+1} \\ = \underbrace{\left(\widetilde{\mathbb{P}}_d + \widetilde{\mathbb{C}}_{d,n+1}^{e-1} : \widetilde{\boldsymbol{H}}_{n+1}^{e} \otimes \boldsymbol{1}\right)}_{\mathbb{G}_3} : \mathrm{d}\boldsymbol{d}_{n+1}\Delta t \\ \underbrace{\widetilde{M}_{n+1}^{e-1}\widetilde{\boldsymbol{H}}_{n+1}^{eT} : \widetilde{\mathbb{S}}_{n+1}}_{\boldsymbol{G}_4} : \mathrm{d}\ \mathrm{dev}\boldsymbol{\tau}_{n+1} + \underbrace{\left(\widetilde{M}_{n+1}^{e-1} + A_n \frac{\partial \xi_{n+1}}{\partial p_{\tau,n+1}}\right)}_{G_5} \mathrm{d}p_{\tau,n+1} \\ = \underbrace{\left(\widetilde{M}_{n+1}^{e-1}\widetilde{\boldsymbol{H}}_{n+1}^{eT} : \widetilde{\mathbb{P}}_d + \boldsymbol{1}\right)}_{\boldsymbol{G}_6} : \mathrm{d}\boldsymbol{d}_{n+1}\Delta t \end{aligned} \tag{B-35}$$

or,

$$\begin{aligned} \mathbb{G}_1 : \mathrm{d}\ \mathrm{dev}\boldsymbol{\tau}_{n+1} + \boldsymbol{G}_2\ \mathrm{d}p_{\tau,n+1} = \mathbb{G}_3 : \mathrm{d}\boldsymbol{d}_{n+1}\Delta t \\ \boldsymbol{G}_4 : \mathrm{d}\ \mathrm{dev}\boldsymbol{\tau}_{n+1} + G_5\ \mathrm{d}p_{\tau,n+1} = \boldsymbol{G}_6 : \mathrm{d}\boldsymbol{d}_{n+1}\Delta t \end{aligned} \tag{B-36}$$

Here, $\mathbb{G}_1$ and $\mathbb{G}_3$ are fourth-order tensors, $\boldsymbol{G}_2$, $\boldsymbol{G}_4$ and $\boldsymbol{G}_6$ are second-order tensors, and $G_5$ is a scalar. The relation in (B-36) can then be solved for the values of $\mathrm{d}\ \mathrm{dev}\boldsymbol{\tau}_{n+1}$ and $\mathrm{d}p_{\tau,n+1}$ obtaining:

$$\mathrm{d}\,\mathrm{dev}\boldsymbol{\tau}_{n+1} = \underbrace{\left\{ \begin{array}{c} \left(\mathbb{G}_1 - \dfrac{\boldsymbol{G}_2 \otimes \boldsymbol{G}_4}{G_5}\right)^{-1} \left(\mathbb{G}_3 - \dfrac{\boldsymbol{G}_2 \otimes \boldsymbol{G}_6}{G_5}\right) + \\ \mathbf{1} \otimes \dfrac{\boldsymbol{G}_6 - \left(\mathbb{G}_1 - \dfrac{\boldsymbol{G}_2 \otimes \boldsymbol{G}_4}{G_5}\right)^{-1} : \left(\mathbb{G}_3 - \dfrac{\boldsymbol{G}_2 \otimes \boldsymbol{G}_6}{G_5}\right)}{G_5} \end{array} \right\}}_{\tilde{\mathbb{C}}^{ep}_{d,n+1}} : \mathrm{d}\boldsymbol{d}_{n+1}\Delta t$$

$$\mathrm{d}p_{\tau,n+1} = \underbrace{\frac{\boldsymbol{G}_6 - \left(\mathbb{G}_1 - \dfrac{\boldsymbol{G}_2 \otimes \boldsymbol{G}_4}{G_5}\right)^{-1} : \left(\mathbb{G}_3 - \dfrac{\boldsymbol{G}_2 \otimes \boldsymbol{G}_6}{G_5}\right)}{G_5}}_{\widetilde{\boldsymbol{H}}^{ep}_{v,n+1}} : \mathrm{d}\boldsymbol{d}_{n+1}\Delta t \qquad \text{(B-37)}$$

Here, special consideration must be observed while solving (B-36), since it involves fourth-order tensors, second-order tensors and scalar values, and appropriate multiplication operations are used to get the results to consistent order. The full-form of the result is not given here for brevity, but can simply be found by substituting the values of $\mathbb{G}_1$, $\boldsymbol{G}_2$, $\mathbb{G}_3$, $\boldsymbol{G}_4$, $G_5$ and $\boldsymbol{G}_6$ from (B-35) in (B-37).

The values of $\mathrm{d}\,\mathrm{dev}\boldsymbol{\tau}_{n+1}$ and $\mathrm{d}p_{\tau,n+1}$ in (B-37) can then be substituted in (B-28) in terms of $\tilde{\mathbb{C}}^{ep}_{d,n+1}$ and $\widetilde{\boldsymbol{H}}^{ep}_{v,n+1}$ which are the fourth-order deviatoric and second-order deviatoric-isochoric coupling elastoplastic consistent material moduli, respectively.

$$d\boldsymbol{\tau}_{n+1} = \underbrace{\left(\tilde{\mathbb{C}}^{ep}_{d,n+1} + \mathbf{1} \otimes \widetilde{\boldsymbol{H}}^{ep}_{v,n+1}\right)}_{\tilde{\mathbb{C}}^{ep}_{\tau,n+1}} : \mathrm{d}\boldsymbol{d}_{n+1}\Delta t \qquad \text{(B-38)}$$

Here, $\tilde{\mathbb{C}}^{ep}_{\tau,n+1}$ is the required elastoplastic consistent tangent modulus in terms of Kirchhoff stress which can then be transformed back in terms of Cauchy stress using:

$$\tilde{\mathbb{C}}^{ep}_{n+1} = J^{e-1}_{n+1}\tilde{\mathbb{C}}^{ep}_{\tau,n+1} = J^{e-1}_{n+1}\left(\tilde{\mathbb{C}}^{ep}_{d,n+1} + \mathbf{1} \otimes \widetilde{\boldsymbol{H}}^{ep}_{v,n+1}\right) \qquad \text{(B-39)}$$

For the case where the deviatoric and volumetric parts of deformation are not coupled, i.e. cubic crystal symmetry, consistent elastoplastic tangent moduli are simply given by:

$$\tilde{\mathbb{C}}^{ep}_{n+1} = J^{e-1}_{n+1}\left(\tilde{\mathbb{C}}^{e-1}_{d,n+1} + \tilde{\mathbb{S}}_{n+1}\right)^{-1} : \tilde{\mathbb{P}}_d + J^{e-1}_{n+1}\left(\tilde{M}^{e-1}_{n+1} + A_n \frac{\partial \xi_{n+1}}{\partial p_{\tau,n+1}}\right)^{-1} \mathbf{1} \otimes \mathbf{1} \qquad \text{(B-40)}$$

## B.4 Homogenisation scheme

The homogenised response of multiple single crystals (grains) at a given material point may sometimes be required to simulate the behaviour of polycrystalline material. This can be achieved using a mean field hypothesis of the partitioning rule. It is required to relate the microscopic quantities discussed in the formulation, $(\boldsymbol{d}, \boldsymbol{w}, \boldsymbol{\sigma})$, with their macroscopic counterparts $(\boldsymbol{D}_M, \boldsymbol{W}_M, \boldsymbol{\Sigma}_M)$. An extended Taylor hypothesis by Asaro and Needleman [64] is used in this work, which is:

$$\boldsymbol{D}_M = \boldsymbol{d}, \qquad \boldsymbol{W}_M = \boldsymbol{w}, \qquad \boldsymbol{\Sigma}_M = \langle \boldsymbol{\sigma} \rangle \qquad \text{(B-41)}$$

Here, $\langle \blacksquare \rangle$ represents volume averaging of a quantity over all the individual single crystals in an aggregate.

Two levels of homogenisation may be required in this model:

i. the homogenised response of individual crystals having different crystal structure in a bicrystal; for example, alternating lamellae of $\alpha$ (HCP) and $\beta$ (BCC) phases in a grain of $\alpha$-$\beta$ titanium alloy
ii. and/or the homogenised response of an aggregate of multiple grains to simulate the response of a polycrystalline material.

Any one, both or none of the above homogenisations may be required for a certain problem. For the case of simulating void growth in a bicrystal composed of single crystals having different crystal structures, volume fractions of each of the crystal type present in bicrystal, $v_1$ and $v_2$, will be required. This is in addition to all the data already mentioned in the formulation, which will now be required for both the crystal types. After getting the values of $\boldsymbol{\sigma}_{n+1}$ and $\tilde{\mathbb{C}}^{ep}_{n+1}$ for each of the crystals, marked as $(\blacksquare)_1$ and $(\blacksquare)_2$, using (B-22) and (B-39), the homogenised response, marked as $(\blacksquare)_T$ was calculated using:

$$(\boldsymbol{\sigma}_{n+1})_T = v_1(\boldsymbol{\sigma}_{n+1})_1 + v_2(\boldsymbol{\sigma}_{n+1})_2, \qquad \left(\tilde{\mathbb{C}}^{ep}_{n+1}\right)_T = v_1\left(\tilde{\mathbb{C}}^{ep}_{n+1}\right)_1 + v_2\left(\tilde{\mathbb{C}}^{ep}_{n+1}\right)_2 \tag{B-42}$$

The macroscopic value of $\xi$, the non-dimensional strain like parameter representative of normalised void volume fraction is represented as $\xi_M$, of a bicrystal is calculated using:

$$\xi_M = v_1\xi_1 + v_2\xi_2 \tag{B-43}$$

where $\xi_1$ and $\xi_2$ are the values of $\xi$ for each of the two phases of the bicrystal. The condition for switching from simple void growth to void coalescence given in (2-34) will then be based on the value of $\xi_M$. It implies that if the normalised void volume fraction of a bicrystal reaches a threshold value, coalescence will start in both phases. The second type of homogenisation can be carried out in the same way.

## B.5 Flow chart

The formulation presented above is implemented in ABAQUS as a user material subroutine, UMAT [65]. Also, the stress state of the integration point for constant stress triaxiality is controlled using multi-point constraint user subroutine (MPC), with details given in [2]. A summary of the implemented model is given in Table B-1.

*Table B-1: Flow chart of model implementation*

1. Following quantities are given at the start of the time increment:

   $$\boldsymbol{d}_{n+1}, \boldsymbol{w}_{n+1}, \left(\mathrm{dev}\boldsymbol{\epsilon}^e_n, \epsilon^e_{kk,n}, \boldsymbol{R}^e_n, \kappa^\chi_{s,n}\right), \left(\mathbb{C}^e_{d0}, \boldsymbol{H}^e_0, M^e_0\right), \boldsymbol{C}_0, \boldsymbol{Z}^\chi_0 = \boldsymbol{s}^\chi_0 \otimes \boldsymbol{m}^\chi_0$$

2. Following are estimated at the start of iteration:

   viscoplastic solution → $\mathrm{dev}\boldsymbol{\tau}_{n+1}$, only for the first-time increment

   for later time increments → $\mathrm{dev}\boldsymbol{\tau}_{n+1} = \mathrm{dev}\boldsymbol{\sigma}_n$

   for first time increment → $p_{\tau,n+1} = \mathrm{M}^e_0 \Delta t d_{kk,n+1}$

   for later time increment→ $p_{\tau,n+1} = p_{\tau,n}$

   forward Euler approx. → $\kappa^\chi_{s,n+1}$

   exponential map with $\tilde{\Omega}^e_n$ → $\boldsymbol{R}^e_{n+1}$

3. Iterations are carried out in a two-level scheme for computations of $\left(\mathrm{dev}\boldsymbol{\tau}_{n+1}, p_{\tau,n+1}, \boldsymbol{R}^e_{n+1}, \kappa^\chi_{s,n+1}\right)$:

   a. Computation of $\boldsymbol{C}_{n+1}, \Delta\boldsymbol{R}^e_{n+1}$:

   $$\boldsymbol{C}_{n+1} = \boldsymbol{R}^e_{n+1}\boldsymbol{C}_0, \qquad \Delta\boldsymbol{R}^e_{n+1} = \boldsymbol{R}^e_{n+1}\boldsymbol{R}^e_n$$

   b. Rotation of $\left(\mathbb{C}^e_{d0}, \boldsymbol{H}^e_0, \boldsymbol{Z}^\chi_0\right)$ to $\left(\tilde{\mathbb{C}}^e_{d,n+1}, \tilde{\boldsymbol{H}}^e_{n+1}, \tilde{\boldsymbol{Z}}^\chi_{n+1}\right)$:

   $$\tilde{\mathbb{C}}^e_{d,n+1} = (\boldsymbol{C}_{n+1} \otimes \boldsymbol{C}_{n+1}) : \mathbb{C}^e_{d0} : (\boldsymbol{C}_{n+1} \otimes \boldsymbol{C}_{n+1})^T$$

   $$\tilde{\boldsymbol{H}}^e_{n+1} = \boldsymbol{C}_{n+1}\boldsymbol{H}^e_0\boldsymbol{C}^T_{n+1}, \qquad \tilde{\boldsymbol{Z}}^\chi_{n+1} = \boldsymbol{C}_{n+1}\boldsymbol{Z}^\chi_0\boldsymbol{C}^T_{n+1}$$

   c. Computation of deviatoric elastic strains $\mathrm{dev}\boldsymbol{\epsilon}^{e*}_{n+1}, \mathrm{dev}\boldsymbol{\epsilon}^e_{n+1}$:

   $$\mathrm{dev}\boldsymbol{\epsilon}^{e*}_{n+1} = \Delta\boldsymbol{R}^e_{n+1}\mathrm{dev}\boldsymbol{\epsilon}^e_n\Delta\boldsymbol{R}^{eT}_{n+1} + \Delta t \mathrm{dev}\boldsymbol{d}_{n+1}$$

   $$\mathrm{dev}\boldsymbol{\epsilon}^e_{n+1} = \tilde{\mathbb{C}}^{e-1}_{d,n+1} : \left(\mathrm{dev}\boldsymbol{\tau}_{n+1} - \tilde{\boldsymbol{H}}^e_{n+1}\epsilon^e_{kk,n+1}\right)$$

   d. Computation of volumetric elastic strains $\epsilon^{e*}_{kk,n+1}, \epsilon^e_{kk,n+1}$:

   $$\epsilon^{e*}_{kk,n+1} = \epsilon^e_{kk,n} + \Delta t\, \mathrm{tr}(\boldsymbol{d}_{n+1})$$

   $$\epsilon^e_{kk,n+1} = M^{e-1}_0\left(p_{\tau,n+1} - \tilde{\boldsymbol{H}}^e_{n+1} : \mathrm{dev}\boldsymbol{\epsilon}^e_{n+1}\right)$$

   e. 1st level – Computation for a new estimate of $\mathrm{dev}\boldsymbol{\tau}_{n+1}$:

N-R method to solve equation → $\mathrm{dev}\boldsymbol{\tau}_{n+1}$

N-R method to solve the equation → $p_{\tau,n+1}$

f. 2nd level – Computation for a new estimate of $\kappa_{s,n+1}^{\chi}$ and $\boldsymbol{R}_{n+1}^{e}$:

N-R method to solve the equation → $\kappa_{s,n+1}^{\chi}$

Exponential map, equation → $\boldsymbol{R}_{n+1}^{e}$

g. Checking for convergence of the iterative scheme:

Whether the difference in values of $\mathrm{dev}\boldsymbol{\tau}_{n+1}, p_{\tau,n+1}$ and $\kappa_{s,n+1}^{\chi}$ are within tolerance?

NO, go back to step a.

YES, continue

4. Calculation of Cauchy stress $\boldsymbol{\sigma}_{n+1}$ using (B-22).
EXIT